\documentclass{llncs}

\usepackage{ifpdf}
\usepackage{times}

\usepackage{cite}
\usepackage{graphicx}
\DeclareGraphicsExtensions{.pdf,.jpeg,.png}
\usepackage{amsmath,amsfonts,amssymb}
\usepackage{mathrsfs}
\usepackage{dsfont}
\usepackage{amsbsy}
\usepackage{stmaryrd}
\usepackage{algorithmic}
\usepackage[tight,footnotesize]{subfigure}
\usepackage{url}
\usepackage{tikz}
\usetikzlibrary{arrows,positioning,automata,circuits.logic.US,circuits.logic.IEC}

\newcommand{\bbR}{\mathbb{R}}

\renewcommand{\thepage}{}
\addtocounter{page}{0}

\renewcommand{\phi}{\varphi}

\renewcommand{\rm}{\rrbracket}

\begin{document}

\title{A temporal logic approach to modular design of synthetic biological circuits}

\author{Ezio Bartocci\inst{1} \and Luca Bortolussi\inst{2,}\inst{3} \and Laura Nenzi\inst{4}}

\institute{
Vienna University of Technology, Austria
%\email{ezio.bartocci@tuwien.ac.at}
\and
DMG, University of Trieste, Italy
\and
CNR/ISTI, Pisa, Italy
%\email{luca@dmi.units.it}
\and
IMT, Lucca, Italy
%\email{laura.nenzi@imtlucca.it}
}

\maketitle

\begin{abstract}\footnote{The final pubblication is available at \url{link.springer.com}}

We present a new approach for the design of a synthetic biological
circuit whose behaviour is specified in terms of signal temporal
logic (STL) formulae. We first show how to characterise with STL formulae
the input/output behaviour of biological modules miming the classical
logical gates (AND, NOT, OR). Hence, we provide the regions of
the parameter space for which these specifications are satisfied.
Given a STL specification of the target circuit to be designed and
the networks of its constituent components, we propose a
methodology to constrain the behaviour of each module,
then identifying the subset of the parameter space in which those
constraints are satisfied, providing also a measure of the robustness
for the target circuit design. This approach, which leverages recent results on the
quantitative semantics of Signal Temporal Logic, is illustrated by
synthesising  a biological implementation of an half-adder.
 
%Per risparmiare spazio facciamo solo una riga le keywords
%\noindent \textbf{Keywords:} Synthetic Biology, Parameter synthesis, Signal Temporal Logic, %compositionality.

\end{abstract}

\pagenumbering{arabic}
\pagestyle{plain}

% !TEX root =  ModulesSysBio.tex

\section{Introduction}
\label{sec:intro}

%****INTRODUCE THE PROBLEM OF SYNTHETIC BIOLOGY AND DE NOVO SYNTHESIS ****
%****SPEAK ABOUT MODULARITY AND TEMPORAL 
%****DESCRIBE OUR APPROACH ****
%****TALK MORE ABOUT LOGIC GATES ****

 {\it Synthetic Biology}~\cite{Fu2009,Szallasi2006} is an emerging discipline 
that aims at the  rational design of {\it artificial} living systems with a 
predictable behaviour, either by creating new biological entities that do 
not exist in nature or by redesigning the existing ones. Even though important 
technological developments have been achieved in this field, the \emph{de-novo} design 
of biological circuits implementing a desired behaviour results to be a very 
 hard task, especially for large scale networks. 
Biological systems are complex to understand and to be engineered: the non-linear 
nature of interactions  reflects in the emergence of systemic behavioural 
properties, not  directly derivable from the knowledge of the individual parts.
To model and control such systems we need to understand 
the relationships between the emergent behaviour and the topology of such 
complex interactions.  A possible approach is to divide the whole system in 
``subunits'' and to look at the structure of the interactions between them. 
This subdivision is often suggested by the way we describe (the components of) 
those systems.  The idea is that compositionality at the specification level, 
to a certain extent, has to be reflected into compositionality at the behavioural 
level. This should depend on the properties satisfied by a single ``subunit'' 
and on the wiring between them.  This way to approach the study of a system 
is called {\it modularity} and the ``subunits'' of the system are called {\it modules}.
Modularity can be effectively achieved in {\it Synthetic Biology}, combining a 
{\it bottom-up}~\cite{Voy2006} and a {\it top-down}~\cite{vondassow2000}  
methodology. The former consists in the assembling of a 
set of well-characterised modules~\cite{Voy2006} together to build sophisticated 
biological circuits and devices. The latter~\cite{vondassow2000} aims 
to identify and characterise the  possible ``subunits'' and this is 
 also helpful to understand  real biological systems, for example to 
discover unknown structures or behaviours or to better understand 
and test current knowledge.

To unveil the system dynamics, it is important to correlate the 
denotation of a module with some of its specific behaviours, and 
understand how the global properties emerge from these local ones. 
This can be performed better if the emergent behaviours are  
specified in a formal language. 
We consider here a \emph{logical} characterisation in terms of (linear) 
temporal logic formulae. In particular, we focus our attention on genetic 
regulatory circuits, seen as networks of interacting genetic modules 
(each representing, for instance, a logic gate). Each module has a set 
of inputs and outputs (usually transcription factors), and its local 
behaviour is specified by temporal logic properties.

In particular, we characterise the
behaviour of  \emph{logic gates} with the addition of  constraints on 
the response time. Logic gates are  physical devices implementing 
a boolean function and they are the fundamental bricks upon 
which all the other logic circuits, including multiplexers, arithmetic 
logic units, memories and microprocessors, are built.
They are primarily implemented using electronic transistors acting 
as electronic switches. In the last decade, genetic circuits acting as logic gates have been 
successfully identified and synthesised~\cite{Myers2009}. This lead researchers to hope to engineer cells to turn them into miniature computers.
%
% (((((( {\color{red}rephrase}
%: However, in the last decade, the successful identification and
%synthesis of genetic circuits acting as logic gates~\cite{Myers2009} have made 
%researchers hoping to turn in the near future  cells into 
%miniature computers. )))))))))

\begin{figure}[!t]
  \centering
  \includegraphics[width=4.2in]{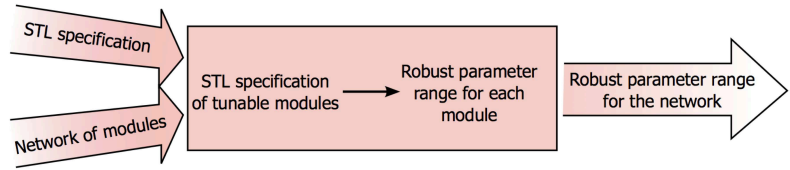}
%\vspace{-8ex}
  \caption{Overview of the proposed approach.}
  \label{fig:overview}
\end{figure}

The main idea of this paper, sketched in Figure~\ref{fig:overview}, is to 
translate the structural compositionality of networks of modules into 
compositionality of local behaviours, exploting it to enforce a set 
of global behaviours to the network. This is realised by identifying a 
subset of parameters for which the truth of local properties implies 
the truth of the global specification, exploiting the modular structure 
of the network. 
We thus interpret  the network of modules as a composition 
of their local properties,  connecting the emergent behaviours with the 
topology of interaction of those local properties.
The technical core  of our approach is the quantitative semantics of 
Signal Temporal Logic~\cite{Maler2004}, which can be seen as a measure 
of robustness of the satisfaction of a certain formula, and which  comes 
with simulation-based methods to compute the robustness score and to 
identify a region of the parameter space in which the formula holds true. 
%In this paper, we concentrate on networks composed of simple logic gates, 
%implementing a boolean function with additional constraints on the response time. 

The contributions of this paper are thus twofold: a design methodology for biological circuits based on a high level logical specification of behaviours and an algorithmic procedure exploiting compositionality to make parameter synthesis more effective, which gives as a byproduct a measure of robustness of the implementation.

%The QMITL-based approach, used to explore the parameter space of modules, can be exploited for the analysis of the global behaviour of more complex networks connecting individual modules. Given a network of modules with their characterising properties, we will determine a subset of the parameter space so to enforce a specific global behaviour, exploiting the modular structure of the network. With this goal in mind,  we interpret  the network of modules as a composition of their properties (local properties) and  we study the relationship between the emergent behaviour and the topology of interaction of those local properties.
%
%In this paper we present a methodology to conduct a quantitative analysis of a network of modules. It is based on Signal Temporal Logic, a dense-time logic with a quantitative semantic that comes with automatic methods to check if a model satisfies a property. The implementation is done in Matlab using Breach, a Matlab/C++ toolbox.

\noindent The paper is structured as follows: in Section \ref{sec:basics} we introduce the background material. In Section~\ref{sec:modules} we discuss the logical characterisation of the basic modules in terms of Signal Temporal Logic (STL).  In Section~\ref{sec:parameter_synthesis} we sketch the algorithmic approach to parameter synthesis and in Section ~\ref{sec:example} we show an application to the design of an half-adder, a fundamental building block of microprocessors. The related works and the final discussion are in Section~\ref{sec:disc}.

% !TEX root =  ModulesSysBio.tex

\section{Background material}
\label{sec:basics}

%{\bf Introduce the model of genetic networks we use. Quickly.
%
%
%Introduce STL and QMiTL. Introduce robustness analysis and Breach.}  

\subsubsection*{Modelling of Gene-Regulatory networks }
In this paper we consider deterministic models of gene regulatory networks, given by a set of non-linear Ordinary Differential Equation (ODE) \cite{Jong2002}. For simplicity, we consider lumped models of gene expression, in which mRNA is not explicitly represented (cf.\ Remark \ref{rem:worstCase} for a further discussion on this point). 
%However, adding it poses no challenges. {\color{red}why adding representation of mRNA does not introduce complications?}.
 We assume to have $n$ genes and proteins. Concentration of protein $i$  at time $t$, $i=1,\ldots,n$,   is denoted by the variable $x_i[t]$, while $\vec{x} = (x_1,\ldots,x_n)$ denotes the vector of concentration variables. The ODE for $x_i[t]$ will then be of the form 
\[ 
\frac {d x_{i}}{dt} = f_{i}(\vec{x}) = f_i^+(\vec{x}) - f_i^-(\vec{x}), 
\] 
where $f_i^+$ is a function giving the net production rate of $x_i$, while  $f_i^-$ is its degradation rate, which is usually a linear function of the form $\mu_i x_i$, for some $\mu_i > 0$.
The function  $f_i^+$, instead, encodes the regulatory mechanism of gene $i$, and is a combination of  Michaelis-Menten or Hill functions \cite{Szallasi2006}.

\subsubsection*{Signal Temporal Logic}
Temporal logic~\cite{Pnueli1977} provides a very elegant framework to specify 
in a compact and formal way an emergent behaviour in terms of {\it time-dependent} events.
Among the myriads of temporal logic extensions available, 
 Signal Temporal Logic~\cite{Maler2004} (STL) is very suitable to characterise 
behavioural patterns in time series of real values generated during the simulation of 
a dynamical system. STL extends the dense-time semantics of Metric Interval Temporal 
Logic~\cite{Alur1996} (MITL), with a set of parametrised numerical predicates playing 
the role of atomic propositions. 
STL provides two different semantics: a boolean semantics that returns yes/no 
depending if the observed trace satisfies or not the STL specification, and
a quantitative semantics that in addition returns  a measure of robustness of the specification.
Recently, Donze et. al~\cite{Donze2013} proposed a very efficient monitoring algorithm for STL 
robustness, now implemented in the Breach~\cite{Donze2010a} tool. 
The combination of robustness and sensitivity-based analysis of STL formulae
 have been successfully applied in several domains, ranging 
from analog circuits~\cite{Jones2010} to systems biology~\cite{Donze2010b,Donze2011}, 
to study the parameter space and also to refine the 
uncertainty of the parameter sets. In the following we recall~\cite{Donze2010} the syntax 
and the quantitative semantics of STL that will be used in the 
rest of the paper. The boolean semantics can be inferred using 
the sign of the quantitative result (positive for true and negative for false).

\begin{definition} [STL syntax] The syntax of the STL is given by
$$\varphi := \top \:|\: \mu \:|\:  \neg \varphi \:|\: \varphi_{1} \wedge \varphi_{2} \:|\: \varphi_{1} \: \mathcal{U}_{[a,b]} \: \varphi_{2}$$
where $\top$ is a true formula, conjunction and negation are the standard boolean
connectives, $[a,b]$ is a dense-time interval with $a<b$ and  $ \mathcal{U}_{[a,b]}$ is the {\it until} operator.

The atomic predicate $\mu :  \mathbb{R}^{n} \rightarrow \mathbb{B}$ is defined as $\mu({\bf x}):= (y({\bf x})\geqslant 0)$, with ${\bf x} [t]=(x_{1}[t],...,x_{n}[t])$, $t \in \mathbb{R}_{\geqslant 0}$, $x_{i} \in \mathbb{R}$,  and $y: \mathbb{R}^{n} \rightarrow \mathbb{R}$ a real-valued function.

\end{definition}

\noindent 
%The time-unbounded operator $\varphi_{1} \: \mathcal{U} \: \varphi_{2}:=\varphi_{1} \: \mathcal{U}_{[0,\infty)} \: \varphi_{2}$  requires $\varphi_{1}$ to hold until $\varphi_{2}$ becomes true.
The (bounded) {\it until} operator $\varphi_{1} \: \mathcal{U}_{[a,b]} \: \varphi_{2}$ requires $\varphi_{1}$ 
to hold from now until, in a time between $a$ and $b$ time units, $\varphi_{2}$ becomes true.
The {\it eventually} operator  $F_{[a,b]}$ and  the {\it always } operator $G_{[a,b]}$ can be defined as usual:
 $F_{[a,b]}  \varphi := \top \mathcal{U}_{[a,b]} \varphi$, $G_{[a,b]} \varphi := \neg F_{[a,b]} \neg \varphi.$

\begin{definition} [{\bf STL  Quantitative Semantics}]
\label{sp.rob}
\begin{align*}
&\rho(\mu, s,t)  & = &\mbox{ } \phantom{a}  y(s[t]) \qquad \mbox{ where } \mu \equiv y(s[t]) \geqslant 0
\\ & \rho (\neg \varphi,s,t)  & =  & \mbox{ } \phantom{a}  - \rho (\varphi,s,t)
\\&\rho( \varphi_{1} \wedge  \varphi_{2}, s,t)  & = &\mbox{ } \phantom{a} \min ( \rho( \varphi_{1},s,t),\rho( \varphi_{2},s,t) )
\\& \rho(  \varphi_{1} \: \mathcal{U}_{[a,b)}  \varphi_{2}, s,t) & = &\mbox{ } \phantom{a}  \max_{t'\in t+[a,b]}(\min(\rho( \varphi_{2},x,t'),\min_{t'' \in [t,t']}(\rho( \varphi_{1},x,t''))))
\end{align*}
where $\rho$ is the quantitative satisfaction function, returning a real number $\rho(\varphi, s,t)$ quantifying the degree of satisfaction of the property $\varphi$ by the signal $s$ at time $t$. Moreover, $\rho(\varphi, s):=\rho(\varphi, s,0)$.
\end{definition}

% !TEX root =  ModulesSysBio.tex

\section{Logical characterisation of modules}
\label{sec:modules}

The approach for the synthesis of 
biological circuits is based on the idea of combining 
simple genetic networks according to a specific design.
These basic building blocks, or modules, are usually 
composed of a single or few genes, and  express a 
specific transcription factor (or signal) in response 
to an input signal, generally  the presence 
or absence of activators or repressors influencing 
the module behaviour. In most of the proposed approaches 
\cite{Terzer2007,Szallasi2006}, such modules are 
the biological equivalent of the logic gates of 
electronics, and as such they encode simple 
boolean functions, like AND, OR, or NOT, that can 
be combined together to build more complex circuits. 
Logic gates are usually described by their truth table. 
However, when moving from electronics to biology, 
the temporal dimension becomes much more relevant, 
and it cannot be  neglected. Furthermore, biological modules 
considered in literature often produce more complex 
input/output (I/O) responses than a boolean I/O relationship, 
like pulses  and oscillations \cite{Szallasi2006}. For this reason, 
we find more convenient to describe the I/O behaviour of a 
module by a set of temporal logic properties. 

%The bottom-up approach to the synthesis of biological circuits is %based on the idea of combining simple genetic networks according %to a specific design. These basic building blocks, or modules, are %usually composed of a single of few genes, and  express a specific %transcription factor (or signal) in response to an input signal, %generally given by the presence or absence of activators or %repressors influencing the module behaviour. In most of proposed %approaches \cite{Terzer2007,Szallasi2006}, such modules are the %biological equivalent of the logic gates of electronics, and as such %they encode simple boolean functions, like AND, OR, or NOT, that %can be combined together to build more complex circuits. Logic %gates are usually described by their truth table. 
%

 More precisely, we define a module $\mathcal{M}$ to be a \emph{genetic network} containing $n$ genes, that produce proteins whose concentration is indicated by $\vec{x}=(x_1,\ldots,x_n)$. The genes of  $\mathcal{M}$ are also regulated by additional $n_I$ external transcription factors, which are the \emph{inputs} of the module. A subset of $n_O$ of the produced proteins constitutes the output of the module. 
The behaviour of such a module is characterised by a set of STL formulae of the form $\phi_I \rightarrow \phi_O$, expressing an I/O relationship, which can be arbitrarily complex. Here $\phi_I$ depends only on the concentration of the input signals $\mathbf{x_I}=(x_{I_{1}},...,x_{I_{n_{I}}})$ and $\phi_O$ only on the concentration of the output signals $\mathbf{x_O}=(x_{O_{1}},...,x_{O_{n_{O}}})$.
Modules can be easily connected into a network, by using one output of a module as the input of another module (see Figure \ref{fig:example}). Such networks can still have external inputs, while a subset of  outputs of their modules will be identified as the output of the network. Furthermore, the network behaviour can also be characterised in terms of a temporal I/O relationship given by STL formulae of the form  $\phi_I \rightarrow \phi_O$. In this sense, a network is nothing but a more complex module, which can then be used as a building block itself, resulting in a hierarchical compositional approach to circuit design. 
%Instead of formally defining modules and networks in general, we discuss here models of  
%In this paper, we consider basic logic gates (AND, OR, and NOT), focussing in particular on how to turn %their truth table, extended with additional temporal information about response time, into an equivalent %set of STL formulae. We will then focus on how to combine such gates into networks implementing more %complex functions, enforcing  specific temporal constraints at the network level.
\subsubsection*{Example: Logic gates.}
As an example, in this paper we consider modules corresponding to AND, OR, and NOT logic gates. For instance, a simple biological implementation of an AND gate can be 
obtained by a module in which a single gene, producing the 
output protein, is activated by two input signals, both required 
to start the gene expression.  This requirement can be enforced 
directly at the level of the gene promoter~\cite{Myers2009} or by letting 
the complex formed by two input proteins activate the gene~\cite{Madec2010}. 
We stick to the first formulation. 
The truth table of the gate is shown in Table \ref{tab:AND}.
To each input and output protein, we associate two thresholds, 
$\theta_{+}$ and $\theta_{-}$. The value \emph{true} in the 
truth table corresponds to a concentration of the 
corresponding protein above  $\theta_{+}$, 
while the value \emph{false} corresponds to the concentration 
being below $\theta_{-}$.  In the truth table we also provide  a 
high level specification 
of the temporal behaviour of the gate, in terms of the 
\emph{maximum response time} $\delta$ and the 
\emph{minimum duration} $\lambda$ of the output signal. 
The former  is an upper bound on the time needed by the 
gate to stabilise. The latter, instead, specifies for how 
long the output remains up or down. This in turn implies 
a constraint on the duration of the input signal: if we want 
the output to remain up for $\lambda$ units of time, then 
both inputs have to remain up for at least $\lambda+\delta$ 
units of time. We can easily turn such a truth table into a 
set of STL formulae,  a formula for each row. 
For instance, the row four of Table 2 gives:
\begin{equation}
\label{eqn:property} 
G_{[0,\lambda+\delta]} (x_A \geq \theta_{A+} \wedge x_B \geq \theta_{B+}) \rightarrow F_{[0,\delta]} G_{[0,\lambda]} (x_C \geq  \theta_{C+}),
\end{equation}
where $x_A$ and $x_B$ are the input signals and $x_C$ is the output. The mathematical model associated with this gate will be given by the non-linear ODE:
\begin{equation}
\label{eqn:ANF_VF}
\dot{x}_C = H_{AND}(x_A,x_B,x_C,\mathbf{k}) = k_{AB} \frac{x_A^n}{K_A^n+x_A^n}\frac{x_B^n}{K_B^n+x_B^n} - k_C x_C, 
\end{equation}
where $\mathbf{k}=(k_{AB}, k_{C}, K_{A}, K_{B}, n)$ is a tuple of 5 parameters: $k_{AB}$, the maximum production 
rate (here we assume a zero basal expression rate), $k_C$, 
the degradation rate, $K_A$ and $K_B$, governing the Hill 
activation function, and $n$, governing the steepness of the 
Hill function. 

%%[Do we want to renormalise $x_i$ into $[0,1]$ and get rid of a parameter?]

The other basic logic gates can be modelled in a similar fashion~\cite{Myers2009}:  
the OR gate can be obtained from the AND gate by a non-collaborative 
activation of gene expression (e.g., replacing in the ODE model the product 
of  Hill functions by a single Hill function depending on the sum of the two concentrations), while the NOT gate can be modeled by a 
gene whose production is repressed by the input protein. For actual 
biological implementations, see for instance the discussion 
in~\cite{Myers2009,Szallasi2006}.

\subsubsection{Example: XOR gate.}
%Having at out disposal AND, OR, and NOT gates, we can easily %construct a XOR gate, as shown in Figure~\ref{fig:example}. 
Figure~\ref{fig:example} shows how to build a XOR gate 
using AND, OR, and NOT gates.
We stress here that the circuit architecture, seen as an implementation of a  boolean function, can be obtained by classical techniques (e.g. by Karnaugh maps~\cite{Karnaugh1953}). 
To fully specify the \emph{extended truth table} of the XOR gate, like for the AND gate (cf. Table \ref{tab:AND}),  we need to specify additional information about the maximum response time and the minimal  duration of the output signal for the network. These two quantities obviously depend on the corresponding ones of the constituent modules. Here we will specify a target temporal behaviour for the network and we will consequently constrain the temporal behaviour of modules.

%
%
%There are two possible to identify such temporal information:  either we fix the temporal behaviour of modules and deduce the one of the network or we specify a target temporal behaviour for the network and we consequently constraint the temporal behaviour of modules. We will pursue this second strategy, as it is more in tune with the idea of design. 

Suppose we fix a maximum response time $\delta$ and a minimum duration $\lambda$ of the output signal for the XOR gate. Looking at Figure \ref{fig:example}, we clearly see that the input signal to the XOR gate has to go through no more than three gates before influencing the output. Hence, if each gate has a maximum response time of $\delta/3$, we obviously obtain a response time for the XOR bounded by $\delta$. To enforce the constraint on the minimum duration of the output signal, we just need to make the output signals of internal gates last sufficiently long to trigger an output signal of the network of the target duration. This can be done by simply taking into account the maximum response delay of each gate.   In the XOR example, we obtain that the AND gates need to have a minimal duration of $\lambda + \delta/3$, while the NOT gates of 
 $\lambda + 2\delta/3$. Clearly, the input signal of the network needs to stay on for $\lambda+\delta$ units of time. 

%
%****SPLIT AND ADD A NEW SUBSUBSECTION, MAKE THIS ARGUMENT SLIGHLY MORE FORMAL (how to compute $\ell$, for instance.****

\subsubsection*{Constraints for arbitrary  \emph{acyclic} networks of logic gates. }
This simple compatibility analysis is easily generalised to arbitrary \emph{acyclic} networks of logic gates, to which we restrict ourselves for the moment. Dealing with feedback loops is more complicated and is left to future investigation. 
%To this end, we aim to exploit STL and its analysis tools based on the quantitative semantics.

Consider a generic module/logic gate in an acyclic network, with target maximum delay $\delta$ and target output signal duration $\lambda$. For each module $\mathcal{M}$ (with a single output) of such a network, let  $\ell_f(\mathcal{M})$ be the length of the \emph{longest} path from $\mathcal{M}$ to an output module (i.e. a module producing one output of the network)
and  $\ell_b(\mathcal{M})$  be the length of the \emph{longest} path from $\mathcal{M}$ to an input module (i.e. a module with an external input). Due to the acyclic nature of the network, both such quantities are finite and can be easily computed by a visit of the graph. Then the processing of an input signal passing from $\mathcal{M}$ has to go through at most $\ell_f(\mathcal{M}) +\ell_b(\mathcal{M}) + 1$ modules, so that a maximum delay of $\delta(\mathcal{M})  = \delta /(\ell_f(\mathcal{M}) +\ell_b(\mathcal{M}) + 1)$ guarantees the response time bound on the network. As for the minimum duration of the output for module $\mathcal{M}$, we can obtain it by the recursive relation $\lambda(\mathcal{M}) = \delta(\mathcal{M}) + \max\{\lambda(\mathcal{M'})\}$, where $(\mathcal{M},\mathcal{M'})$ is an edge of the network, i.e. $\mathcal{M'}$ is a module receiving as input  an output of $\mathcal{M}$. These relationships are easily extended to modules with more than one output, defining a max response time constraint for each output.

We observe here that this compatibility analysis between delays and durations has a counterpart in the STL characterisation of module behaviours. The main idea is that we can express the consistency of the output-input links by the STL formulae like: 
\begin{equation}
\label{eqn:valid}
F_{[\nu_1,\nu_1 + \gamma_1]} G_{[0,\mu_1]} (x \geq  \theta_{+}) \rightarrow 
G_{[\nu_2,\nu_2 + \mu_2]} (x \geq \theta_{+} ),
\end{equation}
This formula states that if a variable is eventually expressed for $\mu_1$ units of time, starting between time $\nu_1$ and $\nu_1 + \gamma_1$, it is for sure expressed for $\mu_2$ units of time, starting at time $\nu_2$. If we set $\mu_1 = \lambda + \delta$, $\mu_2 =  \lambda$, $\gamma_1 = \delta$, and $\nu_2 = \nu_1+\delta$, with $\nu_1\geq 0$, $\lambda,\delta>0$ arbitrary,  we obtain that the formula (\ref{eqn:valid}) is valid. 
According to the previous discussion, we need to choose $\lambda = \lambda(\mathcal{M})$ and $\delta = \delta(\mathcal{M})$. 

\begin{table}[!t]
\begin{center}
\begin{tabular}{ |c|c|c|c| }
\hline
\multicolumn{2}{|c|}{\textbf{Inputs}} & \multicolumn{1}{|c|}{\textbf{Output}} & \textbf{Input$\backslash$Output} \\

\multicolumn{2}{|c|}{max delay=$\delta$} & \multicolumn{1}{|c|}{min. duration=$\lambda$} & \\
\hline
  \textbf{pA} & \textbf{pB} & \textbf{pC} &   \textbf{STL Formula} \\
\hline
  low & low & low & $G_{[0,\lambda+\delta]} (x_A \leq \theta_{A-} \wedge x_B \leq \theta_{B-}) \rightarrow F_{[0,\delta]} G_{[0,\lambda]} (x_C \leq  \theta_{C-})$ \\
\hline
  low & high &  low & $G_{[0,\lambda+\delta]} (x_A \leq \theta_{A-} \wedge x_B \geq \theta_{B+}) \rightarrow F_{[0,\delta]} G_{[0,\lambda]} (x_C \leq  \theta_{C-})$ \\
\hline 
 high & low & low & $G_{[0,\lambda+\delta]} (x_A \geq \theta_{A+} \wedge x_B \leq \theta_{B-}) \rightarrow F_{[0,\delta]} G_{[0,\lambda]} (x_C \leq  \theta_{C-})$\\
\hline 
 high & high & high & $G_{[0,\lambda+\delta]} (x_A \geq \theta_{A+} \wedge x_B \geq \theta_{B+}) \rightarrow F_{[0,\delta]} G_{[0,\lambda]} (x_C \geq  \theta_{C+})$\\
  \hline
\end{tabular}
\end{center}
\caption{Extended truth table for the AND gate}
\label{tab:AND}
\end{table}

\begin{remark}
\label{rem:buildingBlocks}
In principle, we can consider more complex building blocks than logic gates, for instance modules acting as switches or oscillators. To this end, we need to generalise the technique  for combining modules. More specifically, effective connection of modules is enforced by requiring the validity of  formula (\ref{eqn:valid}), which is of the form $\phi_O \rightarrow \phi_I$. Such a formulation in terms of validity of STL formulae can be extended to more general output properties (or proper subformulae thereof). For instance, we can describe oscillations as signals being eventually above a high threshold for some time, and then falling below a low threshold for a subsequent period of time (this property holding globally). 
The subformulae describing these two behaviours can then be matched with input formulae of the kind considered in this paper.

\end{remark}

% !TEX root =  ModulesSysBio.tex

\section{Parameter synthesis}
\label{sec:parameter_synthesis}

%****MODIFY:
%1. chiacchere
%2. definizione formale di worst case signal in termine di STL formulae
%3. TEOREMA per logic gates con dimostrazione sketched
%4 stressare che il teorema � la chiave per param synthesis modulare
%5. commenta su approccio analitico e computazionale
%6. sketch algoritmo*******

Consider a network composed by modules representing logic gates, fix a network specification in terms of an extended truth table/ STL formulae, and consider an ODE model of the network, depending on a tuple of parameters $\mathbf{k}$. We now tackle  the problem of identifying parameters $\mathbf{k}$ such that the network satisfies the specifications. 
%In particular, we want to satisfy both the logical behaviour and the temporal constraints. 
According to the previous section, in order to satisfy the temporal constraints at the network level, we can simply enforce local constraints at the module level. The key intuition of our approach is that modularity can be further exploited, doing parameter synthesis for each module, with a guarantee that the so obtained parametrisation will satisfy the global specification at the network level. 
Furthermore, we will identify a \emph{set} of compatible parameter values rather than a single point. 
%This will immediately give us a measure of the robustness of the network behaviour: the larger such a set, the more robust the system. 
Within the set, furthermore, we can identify an \emph{optimal} parametrisation, by maximising the satisfaction level of the properties, according to STL quantitative semantics. We can also search a biological database, like BioBricks, to find genes with the synthesised  kinetic constraints. 

At the heart of the proposed approach resides the  STL characterization of (the biological implementation of) logic gates.
Essentially, we will restrict to a single gate, fixing the temporal constraints to those implied by the network requirements and by its structure, and find a subset of the parameter space in which the STL formulae characterising the gate behaviour hold true. This can be done algorithmically, using the simulation approach to  parameter synthesis of~\cite{Donze2010b}, based on sensitivity analysis and  STL  quantitative semantics and implemented in Breach~\cite{Donze2010a}. For the simple class of logic gates considered here, we can also do this analytically.
Modularity is the key to the efficiency of our approach: as we treat independently each gate, we just need to explore a low dimensional parameter space, which makes the (computational) procedure feasible. 

\subsubsection*{Modularity of parameter synthesis for logic gates.}
The main difficulty we have to solve is related to the fact that modules are connected in the network, hence they are not independent. Indeed, the expression of a gene is driven by  the dynamical behaviour of its input transcription factors.  The idea to get around this problem is to do a \emph{worst case analysis}, showing that a specific parameter combination satisfies the properties for the ``worst possible input signal'', and that this implies the satisfaction for all possible input signals compatible with the input constraints. This will result in a conservative, but computationally  efficient, estimate. 
%This analysis can be refined by applying the same procedure for a group of gates (a sub-network) rather than for a single gate.
%
We can define the notion of ``worst case input signal''  in terms of the STL characterisation of module behaviour. Given an input signal $\mathbf{x_I}[t]$ of a module $\mathcal{M}$, $t\in[0,T]$, we denote with $\mathbf{x}_{\mathbf{x_I},\mathbf{k}}[t]$ the trajectory of the module, with input $\mathbf{x_I}[t]$ and parameters $\mathbf{k}$.
\begin{definition}
An input signal $\mathbf{\hat{x}_I}[t]$, $t\in [0,T]$ is  a \emph{worst-case input signal}  for the STL specification $\phi_{Input} \rightarrow \phi_{Output}$ of the behaviour of a module $\mathcal{M}$ if and only if, for each parameter configuration $\mathbf{k}$ such that $\rho(\phi_{Input},\mathbf{\hat{x}_I}) \geq 0$ (and $\phi_{Input}$ true) and  $\rho(\phi_{Output},\mathbf{x}_{\mathbf{\hat{x}_I},\mathbf{k}}) > 0$,  the following property holds:
\begin{itemize}
\item for each other input signal $\mathbf{x_I}$ satisfying $\rho(\phi_{Input},\mathbf{x_I}) \geq 0$ (and $\phi_{Input}$ true), it holds that $\rho(\phi_{Output},\mathbf{x}_{\mathbf{x_I},\mathbf{k}}) \geq \rho(\phi_{Output},\mathbf{x}_{\mathbf{\hat{x}_I},\mathbf{k}}) $. 
\end{itemize}
\end{definition}

The characterisation of such a ``worst possible input signal'' depends on the structure of the  target STL formula and on the system of ODE describing a particular module. 
We provide now such a characterisation for the basic logic gate models considered in this paper and for the STL formulae associated with their extended truth tables. 
\\
Consider the property 
$G_{[0,\lambda+\delta]} (x_A \geq \theta_{A^+} \wedge x_B \geq \theta_{B^+}) \rightarrow F_{[0,\delta]} G_{[0,\lambda]} (x_C \geq  \theta_{C^+})$, 
which describes a  row of the extended truth table of an AND gate. This property is of the desired form $\phi_{Input}\rightarrow\phi_{Output}$. Now, $\phi_{Input}$ identifies a subset of trajectories of the space of functions from $[0,\lambda+\delta]$ to $\bbR^2$, i.e. those that satisfy the inequality    $x_A \geq \theta_{A^+} \wedge x_B \geq \theta_{B^+}$ for all $t \in[0,\lambda+\delta]$. Among those functions, we consider $\hat{x}_A[t] \equiv \theta_{A^+}$ and $\hat{x}_B[t]\equiv \theta_{B^+}$, which satisfy $\phi_{Input}$ but have quantitative satisfaction  score equal to zero.
%Clearly, $(\hat{x}_1,\hat{x}_2),0 \models \phi_{Input}$ (and the robustness score $\rho(\phi_{Input},(\hat{x}_1,\hat{x}_2),0) = 0$: we are at the border between satisfaction and non-satisfaction). 
Furthermore, for any other trajectory $x_A[t],x_B[t]$ that satisfies $\phi_{Input}$, we have $x_A[t]\geq \hat{x}_A[t]$ for each $t\in [0,\lambda+\delta]$, and similarly for $x_B$. By monotonicity of Hill functions, this implies that the vector field of the AND gate satisfies
$
f_{AND}(x_A[t],x_B[t],x_C,\mathbf{k}) \geq f_{AND}(\hat{x}_A[t],\hat{x}_B[t],x_C,\mathbf{k})
$
for any $x_C \geq 0$. It then follows, by integrating the vector field, that $x_C[t] \geq \hat{x}_C[t]$ for  $t\in [0,\lambda+\delta]$.
Looking at the satisfaction function of $\phi_{Output}$, defined by
\[
\rho(\phi_{Output},x_C) = \max_{\hat{t}\in[0,\lambda]}( \min_{t \in [\hat{t}, \hat{t}+\delta]}(x_C[t] - \theta_{C^+} )),
\]
it is easy to see that $x_C[t] \geq \hat{x}_C[t]$ for  $t\in [0,\lambda+\delta]$ implies $\rho(\phi_{Output},x_C) \geq \rho(\phi_{Output},\hat{x}_C)$. Hence, any configuration of parameters such that  $\rho(\phi_{Output},\hat{x}_C) > 0$ will imply the truth of $\phi_{Output}$ for any input signal satisfying $\phi_{Input}$, and therefore the truth of $\phi_{Input} \rightarrow \phi_{Output}$. It follows that  $\hat{x}_A,\hat{x}_B$ is a worst-case input signal.

For the AND gate, a similar approach allows us to deal with the other three STL properties associated with the other rows of the truth table. In these cases, we need to find an upper bound for $x_C[t]$, as we need to satisfy the output property $F_{[0,\delta]} G_{[0,\lambda]} (x_C \leq  \theta_{C^-})$. To achieve this, we just need to set $x_J[t]$ to $\theta_{J^-}$, if the input $J$ is false, and to $\gamma_J$ if the input $J$ is true, where $\gamma_{J}$ is the maximum concentration level for the input $x_J$, obtained by dividing maximum production rate by the degradation rate (here $J=A,B$). In fact, in this way we maximise the production rate.
%In the following, we will always rescale the variables $x_i$ by dividing them for $\alpha_i$, so that the maximum concentration level is one. Constants $K$ in the Hill functions need to be rescaled in the same way. 
All this analysis is easily extended to OR and NOT gates, and is captured in the following proposition.
\begin{proposition}
\label{th:worstCase}
Let $x_O$ be the output of a AND or OR logic gate  and let $x_J$ be a generic input. Fix the attention on a row of the extended truth table.
\begin{itemize}
\item If $x_O$ is high, and $x_J$ high, then $\hat{x}_J \equiv \theta_{J^+}$.
\item If $x_O$ is high, and $x_J$ low, then $\hat{x}_J \equiv 0$.
\item If $x_O$ is low, and $x_J$ high, then $\hat{x}_J \equiv \gamma_J$.
\item If $x_O$ is low, and $x_J$ low, then $\hat{x}_J \equiv \theta_{J^-}$.
\end{itemize}
Similarly, let $x_O$ be the output of a NOT logic gate\footnote{The difference between AND/ OR and NOT gates is in the fact that the input is an activator in the first two cases and a repressor in the last one. } and let $x_J$ be its input. Then
\begin{itemize}
\item If $x_O$ is high, then $x_J$ is low and  $\hat{x}_J \equiv \theta_{J^-}$.
\item If $x_O$ is low, then $x_J$ is high and $\hat{x}_J \equiv \theta_{J^+}$.
\end{itemize}
\end{proposition} 
We stress that  this proposition not only allows us to do parameter synthesis modularly, but also to \emph{find a lower bound on the robustness score} of each parameterization.

\begin{remark}
\label{rem:worstCase}
The worst case analysis presented in this section relies on  the monotonicity of the robustness score with respect to the input signal. This follows from the monotone dependence of the output on the input (in fact, $\frac{\partial f}{\partial x_J} > 0$), and of the robustness score on the output. The construction of the worst case input is easily generalised to more complex scenarios satisfying a generalised monotonic property of the robustness score, following \cite{monotoneODE}. As an example, consider a model of the gene expression in which the gene produces the mRNA, and mRNA is in turn translated into the protein. In this case,  for an AND gate, we have an ODE for mRNA similar to the one above, namely $\frac{d m_C}{dt} = f_{AND}(x_A,x_B,m_C,\mathbf{k})$, while the ODE for the protein becomes $\frac{dx_C}{dt} = f_C(m_C,x_C,\vec{k}) =  k_t m_C - k_d x_C$, with  $k_t$ the translation constant and $k_d$ the protein degradation constant. The monotonic dependence of the robustness score (when both inputs are on) from inputs essentially follows because a larger input concentration will produce more mRNA, which in turn will result in a higher expression of the protein, giving a larger robustness degree (input/ output properties are the same). 
If such a monotonic dependence fails, determining the worst case input can be more challenging. We will tackle this issue in our future work.
 
\end{remark}

%[Do we need a theorem here? Like: consider a basic gate. Then, for each row of the extended truth table, there are constant input signals that satisfy the input property, such that, if the gate satisfies also the output property for a set of parameters $\mathbf{k}$, then the output remains true when considering any input signal that satisfies the input property, with increased robustness score...]
%

\subsubsection*{Sketch of the algorithm.}
Assuming the temporal constraints on the extended truth tables of modules have been derived from those of the network, the algorithm for parameter synthesis then work as follows: for any module/gate of the network, and any row in the extended truth table, fix the values of input signals to the worst case ones, and then do STL parameter synthesis  to identify a subset of the parameter space in which the STL formula associated with the row is true.  Take the intersection of these sets for each row in the truth table of each module\footnote{We use the convention that parameters not influencing a gate are set to their whole domain by the STL procedure}.

The STL parameter synthesis can be performed applying the sensitivity-based algorithm~\cite{Donze2010b} implemented in the Matlab toolbox Breach~\cite{Donze2010a}. This is a general approach, applicable to any module for which a worst-case input signal has been identified. However, for logic gates AND, OR, and NOT, we can further exploit their simplicity and characterise analytically a subset of parameters for which the STL specification is satisfied. This is due to the fact that, once the input signals are fixed, the non-linear model of the gate reduces to a linear set of ODEs, for which we can compute the solution in closed form. The details of the  computation are reported in the Appendix.

\section{Example: Half-Adder}
\label{sec:example}

%
%*****1 Introduci half adder; 2. introduci il rescaling in [0,1];  3. spiega i delay globale e sui singoli pezzi. 4. fissa soglie e mostra valori per i K, gli n, e le deg rates. Commenta figure, che funziona! Dire qualcosa su ricerca ottimo valore di robustezza*****

\begin{figure}[tb]
\centering
\includegraphics[width=0.9\textwidth]{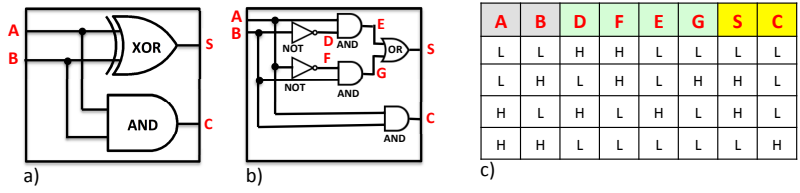}
\caption{a) Half Adder implemented using two logic gates (XOR, AND), b) Half Adder implemented combining  six logic gates, c) truth table for the Half Adder.}
\label{fig:example}
\vspace{-3ex}
\end{figure}

The half-adder is a digital component that perfoms the  sum of two bits A and B 
and provides two outputs, the sum (S) and the carry (C) signal representing 
 an overflow into the next digit of a multi-digit addition. 
The value of the sum is 2C + S. Figure~\ref{fig:example} a) shows the 
simplest half-adder design and it incorporates a XOR gate for
S and an AND gate for C. Figure~\ref{fig:example} b) shows an alternative 
design using two NOT gates,  two AND gates and one OR gate
instead of a XOR gate. This is the design of the half-adder we intend to use, thus exploiting the characterisation of worst-case inputs for AND, OR, and NOT gates given in Proposition \ref{th:worstCase}. Figure~\ref{fig:example} c) shows the output of each component gate of the half-adder, for each pair of  inputs. 

%Two half-adders can be combined in a full-adder as Figure~\ref{fig:example} c)
%shows, with the addition of an OR gate to combine their carry outputs. 
%These adders are the fondamental components in digital electronics
%to build much more sophisticated  Arithmetic Logic Unit (ALU) that
%of the current computer processors.

\begin{figure}[tb]
\hspace{-3.8cm}
\begin{center}
\includegraphics[width=12cm]{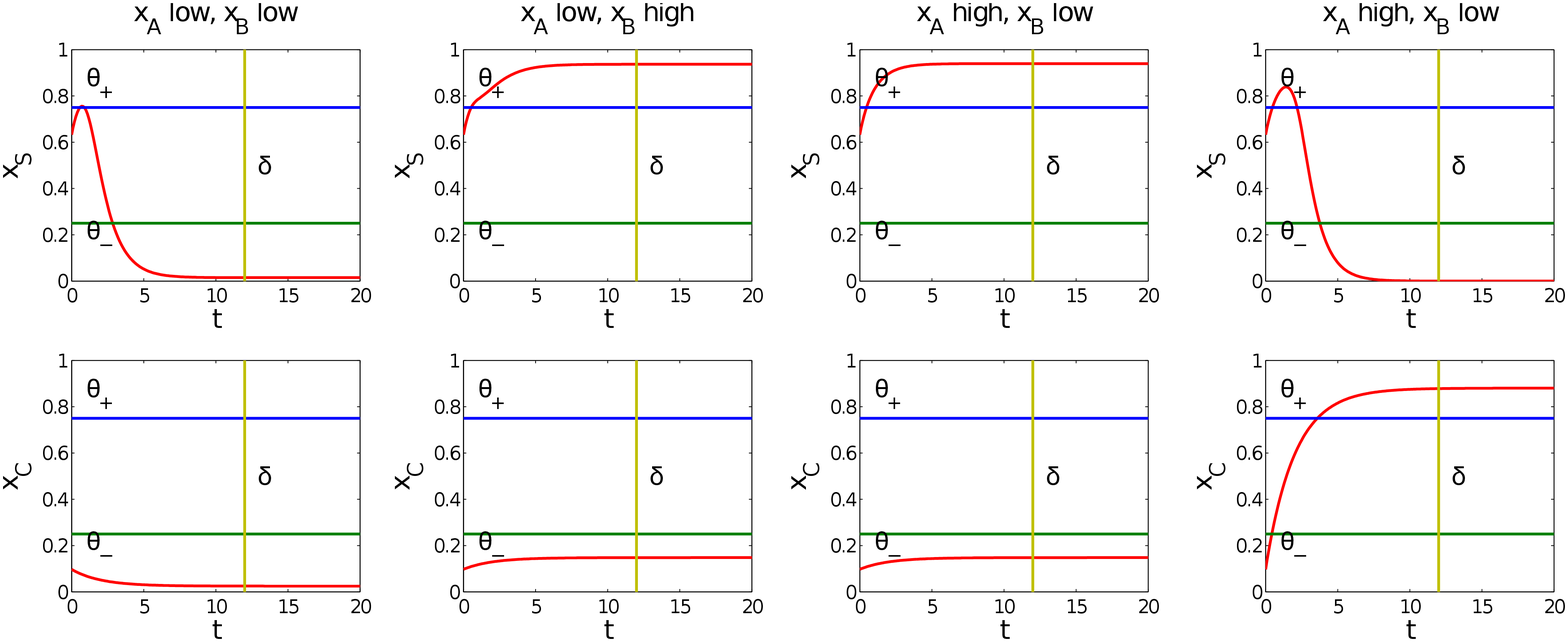}
\end{center}
 \caption{The {\color{red}red} curves represent the output signals of the Half-Adder gate, $S$ and $C$, in the four different combination of the inputs $A$ and $B$, one for each column; the horizontal lines are the threshold concentrations ($\theta_{+}$ in {\color{blue}blue} and $\theta_{-}$ in  {\color{green!60!black}green}); the  {\color{yellow!80!black}yellow} vertical line represents the time bound $\delta$.
}
\label{fig:HADDER}
\vspace{-0.5cm}
\end{figure}

We applied the algorithm discussed in the previous section to such a network layout, fixing the maximum total delay of the half-adder to 12 time units. Applying the method to enforce time constraints to each module, we  obtain that all the gates that are part of the XOR gate must have a maximum time delay of 4 time units, while the AND gate whose output is C can have a maximum response time bounded by 12 time units. Before doing  parameter synthesis, we also rescaled the concentration of each protein to the interval [0,1]. In this way, activation and deactivation thresholds are relative to the maximum steady state expression level of  each protein. For this example,  we then arbitrarily fixed all the activation thresholds to $\theta_+ = 0.75$ and the deactivation thresholds to $\theta_- = 0.25$, and then synthesised set of parameters consistent with the STL network specification and with such thresholds.
We obtained the following bounds for parameters, with indices in the $n$ and $\alpha$  parameters referring to the output variable and indices in the $K$ parameters referring to the input and output protein, as from Figure \ref{fig:example} b). AND gate: $n_C, n_E,n_G \geq 3.2129$, $0.3406 \leq K_{AC}, K_{BC}, K_{AE}, K_{DE}, K_{BG}, K_{FG} \leq 0.4228$, $\alpha_C \geq 0.3074$, $\alpha_E,\alpha_G \geq 0.9222$. OR gate: $n_S \geq 3.1681$, $0.4050 \leq K_{ES}, K_{GS} \leq 0.5090$, $\alpha_S \geq 0.9222$. NOT gates: $n_D, n_F \geq 2.5372$, $0.4192 \leq K_{AF}, K_{BD} \leq 0.4966$, $\alpha_D,\alpha_F \geq 0.9222$. Constraints are similar for all gates of a given class (e.g.\ all AND gates) as a consequence of the rescaling of variables in [0,1]. Obviously, in a further step matching actual biological components to the circuit design, this rescaling  has to be properly accounted for (for instance, by rescaling also the parameters of the biological components).
Picking a value for each parameter consistent with the previous constraints, we can observe in Figure~\ref{fig:HADDER} that the dynamics of the network indeed satisfies the specifications of a half-adder.

 We remark  that, even if in this example we fixed the activation and deactivation thresholds and did parameter synthesis for the other parameters of the model, in the formal derivation we considered such threshold as parameters themselves.

%
%
%Show a model of the half adder and find parameters!

%\input{related}

% !TEX root =  ModulesSysBio.tex

\section{Discussion}
\label{sec:disc}

In this paper we focused on the design techniques for synthetic biological systems. 
We developed an approach based on two ideas: the specification of system properties in terms of signal temporal logic, and the exploitation of modularity to obtain an efficient procedure to identify a set of parameters for which the network satisfies its STL specification. 
In particular, we concentrated on the parameter synthesis problem for \emph{networks of logic gates}, implemented as simple genetic networks. For \emph{acyclic networks}, we are able to identify efficiently a set of parameters satisfying  STL formulae encoding not only the desired boolean behaviour of the network, but also constraints on its response time.

%
%Idea di fondo: design modulare. Qui exploited per paramenter synthesis. 
%Abbiamo considerato logic gates e constraint in termini del comportamento logico E in termini della velocit� di risposta.

Modularity allows us to synthesise parameters efficiently, processing each gate component independently. This is possible by isolating each module from the network assuming the worst possible input, which we  formally characterised for the basic logic gates considered. We then showed the approach at work with a network implementing an half-adder. 

The approach of this paper can be complemented by looking at databases of biological components, like BioBricks \cite{Knight2003}, for actual combinations of gene and promoters that satisfy the constraints on parameters. A delicate point for this plan is that we are implicitly requiring  each module to produce different, non-interfering, output proteins, a not necessarily biologically realistic hypothesis. We will look at possible ways of relaxing this constraint, as in~\cite{Yaman2012}.  
Other directions for future work include the generalisation of Proposition \ref{th:worstCase} to deal with more complex modules, for instance feed-forward networks implementing pulse generation or a low-pass filter. Moreover, we will consider the problem of dealing with more complex network topologies, having feedback loops. We expect to make some progress in this direction by suitably rephrasing parameter synthesis  as the computation of a fixed point.
%, possibly relying on approximate bisimulation techniques XXX. 
Finally, we will also take into account the effects of stochasticity, for instance by exploiting moment closure techniques \cite{vankampen}.

%
%
%qui abbiamo fissato soglie (formule) e cercato parametri. Possiamo fare il viceversa o settare soglie e parametri assieme. 
%
%Limitazioni attuali e future work:
%1. generalizzare il teorema w case a prop pi� complesse: moduli pi� complessi
%2. gestire feedback loops.
%3. interfacciare con biobricks.

\subsubsection*{Related Work.}  
De novo  design of a synthetic biological circuit~\cite{Densmore2009} 
implementing a desired behaviour is a very computational 
 intensive task.  The majority of the existing approaches 
relies on brute-force techniques running sophisticated optimization 
(i.e. evolutionary algorithms~\cite{Francois2004}, 
simulating annealing~\cite{Dasika2008})
algorithms to tune the kinetic parameters~\cite{Terzer2007,Chen2011,Rodrigo2012}  
values in order to match the desired beahaviour. 

These methodologies, lacking of compositionality,
do not scale well and they are very computationally expensive 
for large networks. A more rational approach
for automatic design was proposed by Marchisio and 
Stelling in~\cite{Marchisio2011,Beal2012} where they show
a workflow design taking as input a truth table 
and generating as output several possible circuit 
schemes, ranking them in the order of complexity.
The choice of a truth table as a input specification 
for the target circuit design may be not enough 
when we need to guarantee that the result 
is produced after a proper delay. Additionally, the
design needs to take in consideration the signal compatibility
among the ``wired'' devices (a problem treated 
in~\cite{Yaman2012}): the output signal  
of one device must match (in terms of low/high thresholds) 
with the input signal the other design.  The novelty of our contribution is using signal temporal 
logic as specification language both for the target circuit 
and for the available components, adding also time 
constraints in the design process. 
Furthermore, the device compatibility  is 
rephrased in terms of a STL formula, of the form $\phi_O\rightarrow \phi_I$, 
and the correct matching is elegantly obtained by requiring this formula to be valid.

Another related approach, is the one proposed 
by Batt et al. in~\cite{Batt2007}, where the authors  approximate 
the behaviour of genetic regulatory networks 
with piecewise multi-affine systems. In this class of models, the state-space 
is partitioned in hyper-rectangles exhibiting useful 
convexity properties~\cite{BH06} that allows to compute an 
over-approximation of the reachable sets. 
The authors exploit this characteristic to guide the 
parameter space partitioning in search 
 of the intervals for which 
the gene networks  is enforced to satisfy a particular
 behaviour expressed in a linear temporal logic formula. 
However, their approach is not modular, 
and only the rates of production and degradation of the 
proteins can be chosen as possible parameters.  
Furthermore, by using an over-approximation, 
the property usually expresses invariants and 
the parameter ranges found are very coarse,
without discriminating trajectories with different
time-constraints.

Finally, among the vast literature on combinatorial circuit design, we mention~\cite{maler2004timing}, where authors study the timing behaviour of a acyclic circuits by means of timed automata. Our approach is simpler and motivated by the inherent precision of delays in ODE models. However, the techniques of~\cite{maler2004timing}  could be helpful to relax the timing constraints we impose and to deal with intrinsic variability of biochemical systems.

%*****************************************************************************
%\section*{Acknowledgment}

%*****************************************************************************

\bibliographystyle{plain}
\bibliography{mod}

\clearpage
\appendix
% !TEX root =  ModulesSysBio.tex

\section{Author's contributions} 
L. Nenzi (PhD  student at IMT, Lucca) developed the mathematical and the computational part. 
All authors contributed to brainstorning and to the writing.

\section{Half-Adder, system of ODEs} 
\label{app:ha}

The full ODE system for the Half-Adder  model is:
$$
 \begin{cases}
 \frac {dx_{D}}{dt}=\frac{\beta_{D}}{1+(\frac{x_{B}}{K_{BD}})^{n}} - \alpha_{D} \cdot x_{D},\\ 
\frac {dx_{E}}{dt}=\beta_{E} \cdot \frac {x_{A}^{n}}{K_{AE}^{n}+ x_{A}^{n}} \cdot \frac{x_{D}^{n}}{K_{DE}^{n}+ x_{D}^{n}} - \alpha_{E} \cdot x_{E}, \\
\frac {dx_{F}}{dt}=\frac{\beta_{F}}{1+(\frac{x_{A}}{K_{AF}})^{n}} - \alpha_{F} \cdot x_{F}, \\
\frac {dx_{G}}{dt}=\beta_{G} \cdot \frac{x_{F}^{n}}{K_{FG}^{n}+ x_{F}^{n}}\frac{x_{B}^{n}}{K_{BG}^{n}+ x_{B}^{n}}- \alpha_{G} \cdot x_{G}, \\
\frac {dx_{S}}{dt}=\beta_{S} \cdot  \frac{(\frac{x_{E}}{K_{ES}})^{n}+ (\frac{x_{G}}{K_{GS}})^{n}}{1+(\frac{x_{E}}{K_{ES}})^{n}+(\frac{x_{G}}{K_{GS}})^{n}}- \alpha_{S} \cdot x_{S}, \\
\frac {dx_{C}}{dt} =  \beta_{C} \cdot \frac{ {x_{A}}^{n}}{K_{AC}^{n}+{x_{A}}^{n}}  \cdot  \frac{  {x_{B}}^{n}}{K_{BC}^{n}+{x_{B}}^{n}}  - \alpha_{C} \cdot x_{C}, \\
x_{D}(0)=x_{D_{0}}, \\
x_{E}(0)=x_{E_{0}}, \\
x_{F}(0)=x_{F_{0}}, \\
x_{G}(0)=x_{G_{0}}, \\
x_{S}(0)=x_{S_{0}}, \\
x_{C}(0)=x_{C_{0}};
\end{cases}
$$

where $A$ and $B$ are the inputs of the whole system, $D$ and $F$ are outputs of NOT gates, $E$, $G$ and $C$ are outputs of AND gates and $S$ is the output of an OR gate.

\section{Analytic characterisation of parameter synthesis for logic gates}
\label{app:analytic}

If we fix the value of inputs signals, each gate (AND, NOT, OR) can be described by a linear ODE systems of the form
$$
 \begin{cases}
\frac {dx}{dt} =  \beta \cdot K   - \alpha \cdot x, \\
x(0)=x_0;
\end{cases}
$$
where x is the concentration of the output, $\beta>0$ is the production rate, $\alpha>0$ the degradation rate and $1\geqslant K \geqslant 0$ is the Hill term. We can rescale the systems in $[0, 1]$ observing that, for each $t$, $x(t) \leqslant \frac{\beta}{\alpha}$, the steady state value for $K=1$, provided $x_0  \leqslant \frac{\beta}{\alpha}$. Calling $\gamma= \frac{\beta}{\alpha}$, and $\tilde{x} = x/\gamma$, we have: 

$$
\begin{cases}
\frac {d\tilde{x}}{dt} =\frac {d(\frac{x}{\gamma})}{dt}=  \frac{\beta}{\gamma} \cdot K  - \alpha \cdot \frac{x}{\gamma}= \alpha \cdot K  - \alpha \cdot \tilde{x},\\
\tilde{x}(0)=\tilde{x}_0;
\end{cases}
$$

The analytic solution of this equation, omitting the tilde for simplicity, is: 
\begin{equation}
\label{eqn:solution}
x(t)= K + (x_{0}-  K) \cdot e^{- \alpha \cdot t}
\end{equation}

The constraints on the dynamics expressed by STL formulas, in the simple case of constant inputs, can be translated into a systems of inequalities. 
As an example, consider the AND gate and the STL formula obtained from the first row of the truth table, as discussed in Section \ref{sec:modules}. 
The analytic solution of the AND gate equation, with initial output concentration $x_{C}(0)=0$, (which is a lower bound on any solution with larger initial conditions, hence represents the worst case for the considered scenario) is:
\begin{equation}
\label{eqn:ANDfirst}
x_{C}(t)= \frac{ {x_{A}}^{n}}{K_{AC}^{n}+{x_{A}}^{n}}  \cdot  \frac{  {x_{B}}^{n}}{K_{BC}^{n}+{x_{B}}^{n}} (1- e^{- \alpha \cdot t})
\end{equation}
The STL formula for the fourth row is:
\begin{equation}
\label{eqn:property} 
G_{[0,\lambda+\delta]} (x_A \geq \theta_{A^{+}} \wedge x_B \geq \theta_{B^{+}}) \rightarrow F_{[0,\delta]} G_{[0,\lambda]} (x_C \geq  \theta_{C^{+}}),
\end{equation}
If we fix $x_{A}=\theta_{A^{+}}, x_{B}=\theta_{B^{+}}, x_{C}(0)=0$, the formula is satisfied if and only if
$$
F_{[0,\delta]} G_{[0,\lambda]} (x_C \geq  \theta_{C^{+}}).
$$
Now,  the solution (\ref{eqn:ANDfirst}) is a \emph{monotonic increasing function} converging to the steady state value $x_{C}(\infty)=K$. It follows that if $K \geqslant \theta_{C^{+}}$ and $x(\delta) \geqslant \theta_{C^{+}}$, the STL formula is satisfied (the second condition guarantees that the threshold is crossed no later than $\delta$ time units).
In a similar way it is possible to derive a system of inequalities for all the other STL constraints considered.
\subsubsection*{Bounding the degradation constant.}
We first discuss how to bound the degradation constant. In particular, we will provide a generic bound, holding for all basic logic gates considered.
Fix the thresholds $\theta_{+}$ (high concentration) and $\theta_{-}$ (low concentration) for the output and the maximum time delay $\delta$. We need to consider two cases:
\begin{itemize}
\item Case $x_{0}=0$ and $x(\delta) \geqslant \theta_{{+}}$. Here we want to upper bound by $\delta$ the time at which $x$ crosses the high concentration threshold $\theta_+$.  Now, for the solution $x(t)$ to eventually become bigger than  the threshold $\theta_+$, we need $K>\theta_+$. We can enforce a stricter constraint by setting 
$K \geqslant   \theta_{+} (1+p)$ for $p>0$, which guarantees that the threshold is crossed in finite time.
From equation \ref{eqn:solution} we get
$$
 K(1-e^{-\alpha \delta}) \geqslant \theta_{+}   \qquad \mbox { for } 1 \geqslant K \geqslant   \theta_{+} (1+p),
$$
thus
 $$
\alpha \geqslant \frac{1}{\delta}\log\left(\frac{K}{K-\theta_{+}}\right)  \qquad \mbox { for } 1 \geqslant K \geqslant   \theta_{+} (1+p),
$$
This inequality holds independently of $K$  if and only if:
$$
\alpha \geqslant \frac{1}{\delta}\log\left(\frac{1}{p \theta_{+}}\right)  
$$

\item Case $x_{0}=1$ and $x(\delta) \leqslant \theta_{{-}}$. In this case, we want to upper bound by $\delta$ the time it takes for the solution to fall below the threshold $\theta_{-} < \theta_{+}$. In this case, we require $K\leq (1 - p)  \theta_{-}$, $p>0$, so that  this time is  bounded. From equation (\ref{eqn:solution}), we obtain:
$$
K + (1 -  K) \cdot e^{- \alpha \cdot \delta}\leqslant  \theta_{-}   \qquad \mbox { for } 0 < K \leqslant (1 - p)  \theta_{-} ,
$$
resulting in
$$
\alpha \geqslant \frac{1}{\delta}\log\left(\frac{1-K}{\theta_{-}-K}\right)   \qquad \mbox { for } 0 < K \leqslant (1 - p)  \theta_{-} ,
$$
holding independently of  $K$  if and only if:
$$
\alpha \geqslant \frac{1}{\delta}\log\left(\frac{1}{p\theta_{-}}\right)  
$$
\end{itemize}
As $\theta_{-}<\theta_{+}$, intersecting the two conditions on $\alpha$ we obtain 
\begin{equation}
\label{eqn:alpha}
\alpha \geqslant \frac{1}{\delta}\log\left(\frac{1}{p\theta_{-}}\right)   
\end{equation}

%%%%%    AND
\subsubsection*{AND gate.}

We consider now the constraints specific to an AND gate.
The ODE systems of the AND gate is:
$$
 \begin{cases}
\frac {dx_{C}}{dt} =  \alpha_{C} \cdot \frac{ {x_{A}}^{n}}{K_{AC}^{n}+{x_{A}}^{n}}  \cdot  \frac{  {x_{B}}^{n}}{K_{BC}^{n}+{x_{B}}^{n}}  - \alpha_{C} \cdot x_{C}, \\
x_{C}(0)=x_{C_0};
\end{cases}
$$

where $x_{A}$ and $x_{B}$ are the concentrations of the inputs $A$ and $B$, $K_{AC}$ and  $K_{BC}$ are  the concentration thresholds  of $A$ and $B$ to activate the production of $C$, $n$ is the Hill coefficient. 

According to the discussion of the paper, we will fix the value of $x_A$ and $x_B$ to a constant, either their activation thresholds $\theta_{A^+}$ and $\theta_{B^+}$, or their  deactivation thresholds $\theta_{A^-}$ and $\theta_{B^-}$, or the maximum steady state level $\gamma_A$ and $\gamma_B$. We set
$$K= \frac{ {x_{A}}^{n}}{K_{AC}^{n}+{x_{A}}^{n}}  \cdot  \frac{  {x_{B}}^{n}}{K_{BC}^{n}+{x_{B}}^{n}}.$$

We fix the the output concentration thresholds  $\theta_{C^{+}}$ and $\theta_{C^{-}}$ and the maximum delay time $\delta$. 

Invoking the same argument used for $\alpha$, we will consider new threshold $\tilde{\theta}_{C^{+}}=(1+p)\theta_{C^+}$ and $\tilde{\theta}_{C^{-}}=(1-p)\theta_{C^-}$, and use those to bound the steady state of the ODE system. This guarantees the existence of a lower bound for $\alpha$, independently of $K_{AC}$ and $K_{BC}$.

Now we introduce two methods to find the subspace of the parameters for which the AND gate module satisfies all the four STL formulae, associated with the four rows of the extended truth table. The first method is more intuitive and considers only hypercubic subspaces in the parameter space, at the price of discarding a lot of admissible values. This strong approximation is dropped in the second method, which results to be formally more accurate, but computationally more difficult.
\begin{description}
\item[Method 1:]  We treat the four STL conditions separately.  
\begin{itemize} 
%%%%%  1  AND
\item Case 1 ($x_{A}=\theta_{A^{+}}, x_{B}=\theta_{B^{+}}, x_{C}(0)=0$). Notice that we fix $x_C(0) = 0$ as, by monotonicity of the solution,  the corresponding trajectory is a lower bound on the trajectories starting from $x_C(0) > 0$. In this case, the steady state of the ODE, which is equal to $K$, will be above the activation threshold if and only if
$$
K \geqslant \tilde{\theta}_{C^{+}}.\\
$$
This corresponds to the following condition
$$
 {\theta^{n}_{A^{+}}}\theta^{n}_{B^{+}}- \tilde{\theta}_{C^{+}} (K_{AC}^{n}+{\theta^{n}_{A^{+}}})  \cdot (K_{BC}^{n}+{\theta^{n}_{B^{+}}})\geqslant 0,
$$
which can be rewritten as:
$$
   (K_{AC}^{n}+{\theta^{n}_{A^{+}}})  \cdot (K_{BC}^{n}+{\theta^{n}_{B^{+}}})\leq  
   \frac{\theta^{n}_{A^{+}}}{{\tilde{\theta}_{C^{+}}}^{\frac{1}{2}}} \cdot
   \frac{\theta^{n}_{B^{+}}}{{\tilde{\theta}_{C^{+}}}^{\frac{1}{2}}}.
$$
Now, as all quantities involved are positive, the previous inequality holds if both 
$$
(K_{AC}^{n}+{\theta^{n}_{A^{+}}})  \leq  
   \frac{\theta^{n}_{A^{+}}}{{\tilde{\theta}_{C^{+}}}^{\frac{1}{2}}}
  $$
and
$$
(K_{BC}^{n}+{\theta^{n}_{B^{+}}})\leq  
   \frac{\theta^{n}_{B^{+}}}{{\tilde{\theta}_{C^{+}}}^{\frac{1}{2}}}
$$
are true. We therefore obtain the following conditions on $K_{AC}$ and $K_{BC}$:
$$
\left\{
K_{AC}^{n}\leqslant  \theta^{n}_{A^{+}} (1-{\tilde{\theta}_{C^{+}}}^{\frac{1}{2}})/({\tilde{\theta}_{+}}^{\frac{1}{2}}) ,\quad
 K_{BC}^{n}\leqslant \theta^{n}_{B^{+}} (1-{\tilde{\theta}_{C^{+}}}^{\frac{1}{2}})/({\tilde{\theta}_{+}}^{\frac{1}{2}})
\right\}
$$

%%%%% 2 AND 
\item Case 2 ($x_{A}=\theta_{A^{-}}, x_{B}=\gamma_{B}, x_{C}(0)=1$). In this case, we chose $x_{C}(0)=1$ because this trajectory is an upper bound for all trajectories starting in $x_{C}(0)<1$. We need to impose the condition
$$
K \leqslant \tilde{\theta}_{C^{-}},
$$
which is expanded as
$$
\frac{ {\theta^{n}_{A^{-}}}}{K_{AC}^{n}+{\theta^{n}_{A^{-}}}}  \cdot  \frac{  {\gamma^{n}_{B}}}{K_{BC}^{n}+{\gamma^{n}_{B}}}\leqslant \tilde{\theta}_{C^{-}},
$$
Now, as $\frac{  {\gamma^{n}_{B}}}{K_{BC}^{n}+{\gamma^{n}_{B}}}\leq 1$, the previous condition is satisfied by requiring 
$$
\frac{ {\theta^{n}_{A^{-}}}}{K_{AC}^{n}+{\theta^{n}_{A^{-}}}} \leqslant \tilde{\theta}_{C^{-}},
$$
which turns into the following condition for $K_{AC}$:
$$K_{AC}^{n}\geq  \theta^{n}_{A^{-}} \frac{1-\tilde{\theta}_{C^{-}}}{\tilde{\theta}_{C^{-}}}$$

%%%%% 3 AND
\item Case 3 ($x_{A}=\gamma_{A}, x_{B}=\theta_{B^{-}}, x_{C}(0)=1$). In this case the condition is also 
$K \leqslant \tilde{\theta}_{C^{-}}$. Reasoning symmetrically as in case 2, we then obtain:
$$K_{BC}^{n} \geq  \theta^{n}_{B^{-}} \frac{1-\tilde{\theta}_{^{C-}}}{\tilde{\theta}_{C^{-}}}$$

%%%%% 4 AND
\item Case 4 ($x_{A}=\theta_{A^{-}}, x_{B}=\theta_{B^{-}}, x_{C}(0)=1$). Here we also have to enforce $K \leqslant \tilde{\theta}_{C^{-}}$, which however holds true if the condition for case 2 or that for case 3 holds. 
\end{itemize}

\paragraph{Intersection.} Intersecting the conditions from case 1 to 4, we get the following bounds on $K_{AC}$ and $K_{BC}$:
$$  \theta_{A^{-}} \left( \frac{1 - \tilde{\theta}_{C^{-}}}{\tilde{\theta}_{{C^{-}}} } \right)^{\frac{1}{n}}  \leq K_{AC} \leq \theta_{A^{+}} \left(\frac{ 1 - \tilde{\theta}_{C^{+}}^{\frac{1}{2}}}{\tilde{\theta}_{C^{+}}^{\frac{1}{2}}} \right)^{\frac{1}{n}} $$
$$  \theta_{B^{-}} \left( \frac{1 - \tilde{\theta}_{C^{-}}}{\tilde{\theta}_{{C^{-}}} }\right)^{\frac{1}{n}}  \leq K_{BC} \leq \theta_{B^{+}} \left(  \frac{ 1 - \tilde{\theta}_{C^{+}}^{\frac{1}{2}}}{\tilde{\theta}_{C^{+}}^{\frac{1}{2}}}\right)^{\frac{1}{n}}  $$

The previous constraints are not void if and only if: 
$$  \theta_{A^{-}}^{n} \frac{1 - \tilde{\theta}_{C^{-}}}{\tilde{\theta}_{{C^{-}}} }  \leq \theta_{A^{+}}^{n} \frac{ 1 - \tilde{\theta}_{C^{+}}^{\frac{1}{2}}}{\tilde{\theta}_{C^{+}}^{\frac{1}{2}}} $$ and
$$  \theta_{B^{-}}^{n} \frac{1 - \tilde{\theta}_{C^{-}}}{\tilde{\theta}_{{C^{-}}} }  \leq  \theta_{B^{+}}^{n} \frac{ 1 - \tilde{\theta}_{C^{+}}^{\frac{1}{2}}}{\tilde{\theta}_{C^{+}}^{\frac{1}{2}}}, $$
giving the following constraint on $n$:
$$n \geq \frac{1}{\min\{ \log(\frac{\theta_{B^{+}}}{\theta_{B^{-}}}),\log(\frac{\theta_{A^{+}}}{\theta_{A^{-}}})\}}  
\log\left(\frac{\tilde{\theta}_{C^{+}}^{\frac{1}{2}}}{\tilde{\theta}_{{C^{-}}}}  \cdot \frac{1 - \tilde{\theta}_{{C^{-}}}}{1 - \tilde{\theta}_{C^{+}}^{\frac{1}{2}}}\right). $$ 

\item[Method 2:]  We provide now more precise bounds for $K_{AC}$ and $K_{BC}$.   \\
\begin{itemize}
%%%%%  1  AND
\item Case 1 ($x_{A}=\theta_{A^{+}}, x_{B}=\theta_{B^{+}}, x_{C}(0)=0$). We study the inequality:
$$
K \geqslant \tilde{\theta}_{C^{+}},\\
$$
that is
$$
\frac{ {\theta^{n}_{A^{+}}}}{K_{AC}^{n}+{\theta^{n}_{A^{+}}}}  \cdot  \frac{  {\theta^{n}_{B^{+}}}}{K_{BC}^{n}+{\theta^{n}_{B^{+}}}}\geqslant \tilde{\theta}_{C^{+}},
$$

Note that, since  $ \frac{ {\theta^{n}_{A^{+}}}}{K_{AC}^{n}+{\theta^{n}_{A^{+}}}} \leqslant 1$ and   $\frac{  {\theta^{n}_{B^{+}}}}{K_{BC}^{n}+{\theta^{n}_{B^{+}}}} \leqslant 1$, there exists a solution if and only if:
$$
\Big\{  \frac{ {\theta^{n}_{A^{+}}}}{K_{AC}^{n}+{\theta^{n}_{A^{+}}}}  \geqslant \tilde{\theta}_{C^{+}},  \frac{  {\theta^{n}_{B^{+}}}}{K_{BC}^{n}+{\theta^{n}_{B^{+}}}}\geqslant \tilde{\theta}_{C^{+}} \Big\},$$ i.e. if and only if
 $$
\Big\{ 0 \leqslant K_{AC}\leqslant  \theta_{A^{+}} \Big( \frac{1-\tilde{\theta}_{C^{+}}}{\tilde{\theta}_{C^{+}}}\Big)^{\frac{1}{n}},  0\leqslant K_{BC}\leqslant \theta_{B^{+}} \Big(\frac{1-\tilde{\theta}_{C^{+}}}{\tilde{\theta}_{C^{+}}}\Big)^{\frac{1}{n}} \Big\}$$

Now, within this rectangle, we need to restrict to the region below the curve 
$$K_{AC}=\theta_{A^{+}} \Big(\frac{\theta^{n}_{B^{+}}}{\tilde{\theta}_{C^{+}}(K_{BC}^{n}+\theta^{n}_{B^{+}})}-1\Big)^{\frac{1}{n}} .$$

Hence, the set of parameters satisfying case 1 is characterised by

\begin{eqnarray*}
& \Big\{ K_{AC}\leqslant  \theta_{A^{+}} \Big(\frac{1-\tilde{\theta}_{C^{+}}}{\tilde{\theta}_{C^{+}}}\Big)^{\frac{1}{n}},  K_{BC}\leqslant \theta_{B^{+}} \Big(\frac{1-\tilde{\theta}_{C^{+}}}{\tilde{\theta}_{C^{+}}}\Big)^{\frac{1}{n}}\Big\}\cap &  \\
& \cap\Big\{ K_{AC}^{n}\leqslant \theta_{A^{+}} \Big(\frac{\theta^{n}_{B^{+}}}{\tilde{\theta}_{C^{+}}(K_{BC}^{n}+\theta^{n}_{B^{+}})}-1\Big)^{\frac{1}{n}} \Big\}&
\end{eqnarray*}

%%%%% 2 AND 
\item Case 2 ($x_{A}=\theta_{A^{-}}, x_{B}=\gamma_{B}, x_{C}(0)=1$): We have to enforce the inequality: 
$$
K \leqslant \tilde{\theta}_{C^{-}},
$$
i.e.
$$
\frac{ {\theta^{n}_{A^{-}}}}{K_{AC}^{n}+{\theta^{n}_{A^{-}}}}  \cdot  \frac{  {\gamma^{n}_{B}}}{K_{BC}^{n}+{\gamma^{n}_{B}}}\leqslant \tilde{\theta}_{C^{-}},
$$

First note that because $ \frac{ {\theta^{n}_{A^{-}}}}{K_{AC}^{n}+{\theta^{n}_{A^{-}}}}  \leqslant 1$ and $\frac{  {\gamma^{n}_{B}}}{K_{BC}^{n}+{\gamma^{n}_{B}}}\leqslant  1$, the truth of 
if $ \frac{ {\theta^{n}_{A^{-}}}}{K_{AC}^{n}+{\theta^{n}_{A^{-}}}} \leqslant \tilde{\theta}_{C^{-}} $ or $\frac{  {\gamma^{n}_{B}}}{K_{BC}^{n}+{\gamma^{n}_{B}}}\leqslant \tilde{\theta}_{C^{-}}$ implies the satisfaction of the target inequality. Therefore

$$
\Big\{ K_{AC}\geqslant  \theta_{A^{-}} \Big(\frac{1-\tilde{\theta}_{C^{-}}}{\tilde{\theta}_{C^{-}}}\Big)^{\frac{1}{n}}  \Big\} 
 \cup 
  \Big \{ K_{BC}\geqslant \gamma_{B} \Big(\frac{1-\tilde{\theta}_{C^{-}}}{\tilde{\theta}_{C^{-}}} \Big)^{\frac{1}{n}}\Big \}  
$$
is a subspace of the parameter space in which the inequality is true.

In the remaining subspace $$
\Big\{ K_{AC}< \theta_{A^{-}} \Big(\frac{1-\tilde{\theta}_{C^{-}}}{\tilde{\theta}_{C^{-}}}\Big)^{\frac{1}{n}}, K_{BC}< \gamma_{B} \Big(\frac{1-\tilde{\theta}_{C^{-}}}{\tilde{\theta}_{C^{-}}} \Big)^{\frac{1}{n}}\Big \}, 
$$
we need to restrict to the region above the curve
$$
K_{AC}=\theta_{A^{-}} \Big(\frac{\gamma^{n}_{B}}{\tilde{\theta}_{C^{-}}(K_{BC}^{n}+\gamma^{n}_{B})}-1\Big)^{\frac{1}{n}}. $$
Hence, the set of parameters satisfying case 2 is

$$
\Big\{ K_{AC}\geqslant  \theta_{A^{-}} \Big(\frac{1-\tilde{\theta}_{C^{-}}}{\tilde{\theta}_{C^{-}}}\Big)^{\frac{1}{n}}  \Big\} 
 \cup 
  \Big \{ K_{BC}\geqslant \gamma_{B} \Big(\frac{1-\tilde{\theta}_{C^{-}}}{\tilde{\theta}_{C^{-}}} \Big)^{\frac{1}{n}}\Big \}  \cup
$$
$$
\cup  \Big\{ K_{AC}< \theta_{A^{-}} \Big(\frac{1-\tilde{\theta}_{C^{-}}}{\tilde{\theta}_{C^{-}}}\Big)^{\frac{1}{n}}, K_{BC}< \gamma_{B} \Big(\frac{1-\tilde{\theta}_{C^{-}}}{\tilde{\theta}_{C^{-}}} \Big)^{\frac{1}{n}}, 
$$
$$
K_{AC}\geqslant \theta_{A^{-}} \Big(\frac{\gamma^{n}_{B}}{\tilde{\theta}_{C^{-}}(K_{BC}^{n}+\gamma^{n}_{B})}-1\Big)^{\frac{1}{n}}\Big \}, 
$$

%%%%% 3 AND
\item Case 3 ($x_{A}=\gamma_{A}, x_{B}=\theta_{B^{-}}, x_{C}(0)=1$): this case is symmetric to case 2, just switching the role of input variables. 
%$$
%K \leqslant \tilde{\theta}_{C^{-}},
%$$
%i.e:
%$$
%\frac{ {\theta^{n}_{A^{+}}}}{K_{AC}^{n}+{\theta^{n}_{A^{+}}}}  \cdot  \frac{  {\theta^{n}_{B^{-}}}}{K_{BC}^{n}+{\theta^{n}_{B^{-}}}}\leqslant \tilde{\theta}_{C^{-}}.
%$$
%
We then obtain the following set of parameters

$$
\Big\{ K_{AC}\geqslant  \gamma_{A} \Big(\frac{1-\tilde{\theta}_{C^{-}}}{\tilde{\theta}_{C^{-}}}\Big)^{\frac{1}{n}}  \Big\} 
 \cup 
  \Big \{ K_{BC}\geqslant \theta_{B^{-}} \Big(\frac{1-\tilde{\theta}_{C^{-}}}{\tilde{\theta}_{C^{-}}} \Big)^{\frac{1}{n}}\Big \}  \cup
$$
$$
\cup  \Big\{ K_{AC}< \gamma_{A} \Big(\frac{1-\tilde{\theta}_{C^{-}}}{\tilde{\theta}_{C^{-}}}\Big)^{\frac{1}{n}}, K_{BC}< \theta_{B^{-}} \Big(\frac{1-\tilde{\theta}_{C^{-}}}{\tilde{\theta}_{C^{-}}} \Big)^{\frac{1}{n}}, 
$$
$$
K_{AC}\geqslant \gamma_{A} \Big(\frac{\theta^{n}_{B^{-}}}{\tilde{\theta}_{C^{-}}(K_{BC}^{n}+\theta^{n}_{B^{-}})}-1\Big)^{\frac{1}{n}}\Big \}, 
$$

%%%%% 4 AND
\item Case 4 ($x_{A}=\theta_{A^{-}}, x_{B}=\theta_{B^{-}}, x_{C}(0)=1$):  A similar argument to case 2 can be used here to obtain the following parameter set

%
%The inequality is again:
%$$
%K \leqslant \theta_{C^{-}},
%$$
%then:
%$$
%\frac{ {\theta^{n}_{A^{-}}}}{K_{AC}^{n}+{\theta^{n}_{A^{-}}}}  \cdot  \frac{  {\theta^{n}_{B^{-}}}}{K_{BC}^{n}+{\theta^{n}_{B^{-}}}}\leqslant \theta_{C^{-}},
%$$
%
%and the set of parameters satisfying case 4 is:

$$
\Big\{ K_{AC}\geqslant  \theta_{A^{-}} \Big(\frac{1-\tilde{\theta}_{C^{-}}}{\tilde{\theta}_{C^{-}}}\Big)^{\frac{1}{n}}  \Big\} 
 \cup 
  \Big \{ K_{BC}\geqslant \theta_{B^{-}} \Big(\frac{1-\tilde{\theta}_{C^{-}}}{\tilde{\theta}_{C^{-}}} \Big)^{\frac{1}{n}}\Big \}  \cup
$$
$$
\cup  \Big\{ K_{AC}< \theta_{A^{-}} \Big(\frac{1-\tilde{\theta}_{C^{-}}}{\tilde{\theta}_{C^{-}}}\Big)^{\frac{1}{n}}, K_{BC}< \theta_{B^{-}} \Big(\frac{1-\tilde{\theta}_{C^{-}}}{\tilde{\theta}_{C^{-}}} \Big)^{\frac{1}{n}}, 
$$
$$
K_{AC}\geqslant \theta_{A^{-}} \Big(\frac{\theta^{n}_{B^{-}}}{\tilde{\theta}_{C^{-}}(K_{BC}^{n}+\theta^{n}_{B^{-}})}-1\Big)^{\frac{1}{n}}\Big \}, 
$$

\end{itemize}

\paragraph{Intersection.}

The intersection of the conditions of cases 2,3 and 4 gives:
$$
\Big\{ K_{AC}\geqslant  \theta_{A^{-}} \Big(\frac{1-\tilde{\theta}_{C^{-}}}{\tilde{\theta}_{C^{-}}}\Big)^{\frac{1}{n}}, 
 K_{BC}\geqslant \theta_{B^{-}} \Big(\frac{1-\tilde{\theta}_{C^{-}}}{\tilde{\theta}_{C^{-}}} \Big)^{\frac{1}{n}}\Big \}  \bigcup
$$
$$
\bigcup  \Big\{ K_{BC}< \theta_{B^{-}} \Big(\frac{1-\tilde{\theta}_{C^{-}}}{\tilde{\theta}_{C^{-}}} \Big)^{\frac{1}{n}}, 
$$
$$
K_{AC}\geqslant \max \Big( \theta_{A^{-}} \Big(\frac{\gamma^{n}_{B}}{\tilde{\theta}_{C^{-}}(K_{BC}^{n}+\gamma^{n}_{B})}-1\Big)^{\frac{1}{n}}, \gamma_{A} \Big(\frac{\theta^{n}_{B^{-}}}{\tilde{\theta}_{C^{-}}(K_{BC}^{n}+\theta^{n}_{B^{-}})}-1\Big)^{\frac{1}{n}} \Big) \Big\} \bigcup 
$$
$$
\bigcup  \Big\{ K_{BC}\geqslant \theta_{B^{-}} \Big(\frac{1-\tilde{\theta}_{C^{-}}}{\tilde{\theta}_{C^{-}}} \Big)^{\frac{1}{n}}, K_{AC}< \theta_{A^{-}} \Big(\frac{1-\tilde{\theta}_{C^{-}}}{\tilde{\theta}_{C^{-}}}\Big)^{\frac{1}{n}},
$$
$$
K_{AC}\geqslant \theta_{A^{-}} \Big(\frac{\gamma^{n}_{B}}{\tilde{\theta}_{C^{-}}(K_{BC}^{n}+\gamma^{n}_{B})}-1\Big)^{\frac{1}{n}} \Big\}, 
$$

Taking the intersection with the condition of case 1 finally gives:
\begin{eqnarray*}
&\Big\{  \theta_{A^{-}} \Big(\frac{1-\tilde{\theta}_{C^{-}}}{\tilde{\theta}_{C^{-}}}\Big)^{\frac{1}{n}} \leqslant  K_{AC} \leqslant  \theta_{A^{+}} \Big(\frac{1-\tilde{\theta}_{C^{+}}}{\tilde{\theta}_{C^{+}}}\Big)^{\frac{1}{n}},&\\
 &\theta_{B^{-}} \Big(\frac{1-\tilde{\theta}_{C^{-}}}{\tilde{\theta}_{C^{-}}} \Big)^{\frac{1}{n}} \leqslant K_{BC} \leqslant \theta_{B^{+}} \Big(\frac{1-\tilde{\theta}_{C^{+}}}{\tilde{\theta}_{C^{+}}}\Big)^{\frac{1}{n}},&\\
& K_{AC}\leqslant \theta_{A^{+}} \Big(\frac{\theta^{n}_{B^{+}}}{\tilde{\theta}_{C^{+}}(K_{BC}^{n}+\theta^{n}_{B^{+}})}-1\Big)^{\frac{1}{n}} \Big\}  \bigcup& \\
&\bigcup  \Big\{ K_{BC}< \theta_{B^{-}} \Big(\frac{1-\tilde{\theta}_{C^{-}}}{\tilde{\theta}_{C^{-}}} \Big)^{\frac{1}{n}}, K_{AC} \leqslant  \theta_{A^{+}} \Big(\frac{1-\tilde{\theta}_{C^{+}}}{\tilde{\theta}_{C^{+}}}\Big)^{\frac{1}{n}},&\\
& \max \Big( \theta_{A^{-}} \Big(\frac{\gamma^{n}_{B}}{\tilde{\theta}_{C^{-}}(K_{BC}^{n}+\gamma^{n}_{B})}-1\Big)^{\frac{1}{n}}, \gamma_{A} \Big(\frac{\theta^{n}_{B^{-}}}{\tilde{\theta}_{C^{-}}(K_{BC}^{n}+\theta^{n}_{B^{-}})}-1\Big)^{\frac{1}{n}} \Big) \leqslant K_{AC}, &\\
& K_{AC}  \leqslant  \theta_{A^{+}} \Big(\frac{\theta^{n}_{B^{+}}}{\tilde{\theta}_{C^{+}}(K_{BC}^{n}+\theta^{n}_{B^{+}})}-1\Big)^{\frac{1}{n}}\Big\} \bigcup & \\
&\bigcup  \Big\{ K_{BC}\geqslant \theta_{B^{-}} \Big(\frac{1-\tilde{\theta}_{C^{-}}}{\tilde{\theta}_{C^{-}}} \Big)^{\frac{1}{n}}, K_{AC}< \theta_{A^{-}} \Big(\frac{1-\tilde{\theta}_{C^{-}}}{\tilde{\theta}_{C^{-}}}\Big)^{\frac{1}{n}},  & \\
& \theta_{A^{-}} \Big(\frac{\gamma^{n}_{B}}{\tilde{\theta}_{C^{-}}(K_{BC}^{n}+\gamma^{n}_{B})}-1\Big)^{\frac{1}{n}} \leqslant K_{AC} \leqslant  \theta_{A^{+}} \Big(\frac{\theta^{n}_{B^{+}}}{\tilde{\theta}_{C^{+}}(K_{BC}^{n}+\theta^{n}_{B^{+}})}-1\Big)^{\frac{1}{n}}  \Big\}. & \\
\end{eqnarray*}
A better understanding of this set can be obtained by inspecting Figures \ref{pAND1} and \ref{pAND2}.

Finally, we can deduce constraints on the  parameter $n$.  For the previous set to be non-void, we need to require
$$  \theta_{A^{-}}^{n} \frac{1 - \tilde{\theta}_{C^{-}}}{\tilde{\theta}_{{C^{-}}} }  \leq \theta_{A^{+}}^{n} \frac{ 1 - \tilde{\theta}_{C^{+}}}{\tilde{\theta}_{C^{+}}} $$ and
$$  \theta_{B^{-}}^{n} \frac{1 - \tilde{\theta}_{C^{-}}}{\tilde{\theta}_{{C^{-}}} }  \leq  \theta_{B^{+}}^{n} \frac{ 1 - \tilde{\theta}_{C^{+}}}{\tilde{\theta}_{C^{+}}}, $$
resulting in
$$n \geq \frac{1}{\min\{ \log(\frac{\theta_{B^{+}}}{\theta_{B^{-}}}),\log(\frac{\theta_{A^{+}}}{\theta_{A^{-}}})\}}  
\log\left(\frac{\tilde{\theta}_{C^{+}}}{\tilde{\theta}_{{C^{-}}}}  \cdot \frac{1 - \tilde{\theta}_{{C^{-}}}}{1 - \tilde{\theta}_{C^{+}}}\right).$$ 

\end{description}

\paragraph{Numerical example.} Let $\theta_+ = 2/3$ and $\theta_- = 1/3$ for all species $A$, $B$, and $C$, $\gamma_{A}=\gamma_{B}=1$ and $p = 0.1$. 
Applying the bounds of the first method, we obtain 
$$n \geq 3.798$$
Then, setting for instance $n=4$, we get
$$0.4120 \leq K_{AC} \leq 0.4267$$
and a similar value for $K_{BC}$.

The second method gives us $$n \geq 2.6818.$$ If we set again $n=4$, for comparison, the subspace of parameters for which the four STL properties are satisfied is given by the region delimited by the three following curves:
$$
 K_{AC}= \theta_{A^{+}} \Big(\frac{\theta^{n}_{B^{+}}}{\tilde{\theta}_{C^{+}}(K_{BC}^{n}+\theta^{n}_{B^{+}})}-1\Big)^{\frac{1}{n}}  
$$
$$
 K_{AC}= \theta_{A^{-}} \Big(\frac{\gamma^{n}_{B}}{\tilde{\theta}_{C^{-}}(K_{BC}^{n}+\gamma^{n}_{B})}-1\Big)^{\frac{1}{n}},   
$$
$$
 K_{AC}=\gamma_{A} \Big(\frac{\theta^{n}_{B^{-}}}{\tilde{\theta}_{C^{-}}(K_{BC}^{n}+\theta^{n}_{B^{-}})}-1\Big)^{\frac{1}{n}} 
$$  
This region  is visually depicted  in Figure \ref{pAND1}. We can observe that the box identified by the first method is strictly included in the set provided by the second method.

 \begin{figure}
\centering
\includegraphics[width=0.9\textwidth]{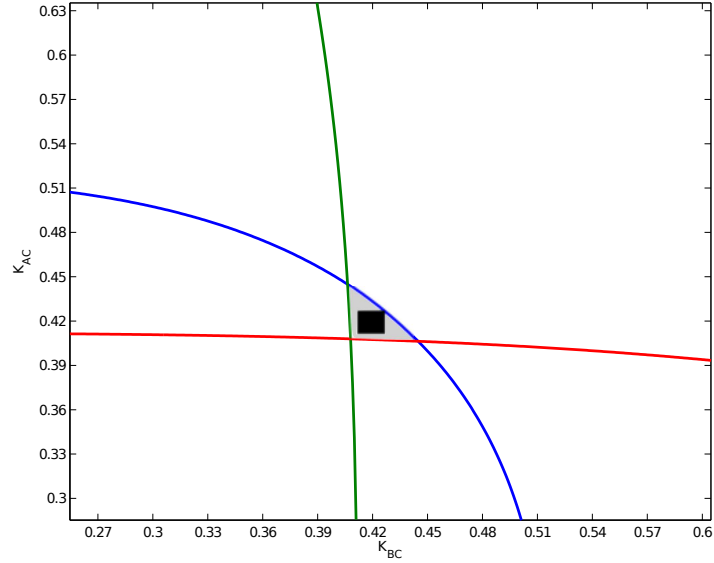}

\caption{The grey region inside the intersection of the curves $
 K_{AC}= \theta_{A^{+}} \Big(\frac{\theta^{n}_{B^{+}}}{\tilde{\theta}_{C^{+}}(K_{BC}^{n}+\theta^{n}_{B^{+}})}-1\Big)^{\frac{1}{n}} 
$(in {\color{blue}blue}),
$
 K_{AC}= \theta_{A^{-}} \Big(\frac{\gamma^{n}_{B}}{\tilde{\theta}_{C^{-}}(K_{BC}^{n}+\gamma^{n}_{B})}-1\Big)^{\frac{1}{n}}
$ (in  {\color{red}red} ) and
$
 K_{AC}=\gamma_{A} \Big(\frac{\theta^{n}_{B^{-}}}{\tilde{\theta}_{C^{-}}(K_{BC}^{n}+\theta^{n}_{B^{-}})}-1\Big)^{\frac{1}{n}} 
$ (in {\color{green!60!black}green})  is the validity domain of the parameters $K_{AC}$ and $K_{BC}$ for   $\theta_+ = 2/3$ and $\theta_- = 1/3$. The black region is the one identified by the first method.}
\label{pAND1}
\end{figure}

If we set  the thresholds to $\theta_+ = 3/4$ and $\theta_- = 1/4$, the first method gives us  $n\geq 3.2129 $, so that for $n=4$, we obtain  $0.3406 \leq K_{AC}, K_{BC} \leq 0.4228$, hence a larger interval.
The validity domain found by the  second approach, instead,  is represented in Figure \ref{pAND2}. Also in this case, the region is larger.
    
 \begin{figure}
\centering
\includegraphics[width=0.9\textwidth]{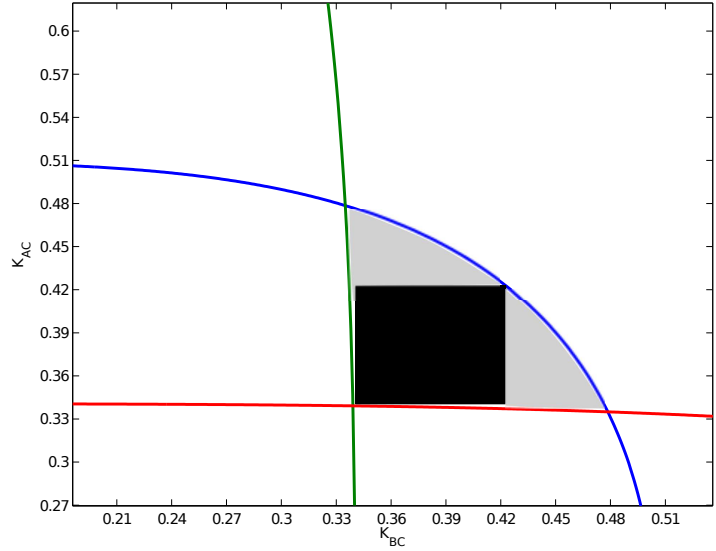}
\caption{The grey region inside the intersection of the curves $
 K_{AC}= \theta_{A^{+}} \Big(\frac{\theta^{n}_{B^{+}}}{\tilde{\theta}_{C^{+}}(K_{BC}^{n}+\theta^{n}_{B^{+}})}-1\Big)^{\frac{1}{n}} 
$(in {\color{blue}blue}),
$
 K_{AC}= \theta_{A^{-}} \Big(\frac{\gamma^{n}_{B}}{\tilde{\theta}_{C^{-}}(K_{BC}^{n}+\gamma^{n}_{B})}-1\Big)^{\frac{1}{n}}
$ (in {\color{red}red}) and
$
 K_{AC}=\gamma_{A} \Big(\frac{\theta^{n}_{B^{-}}}{\tilde{\theta}_{C^{-}}(K_{BC}^{n}+\theta^{n}_{B^{-}})}-1\Big)^{\frac{1}{n}} 
$ (in  {\color{green!60!black}green})  is the validity domain of the parameters $K_{AC}$ and $K_{BC}$ for   $\theta_+ = 3/4$ and $\theta_- = 1/4$. The black region is the one identified by the first method.}
\label{pAND2}
\end{figure}

Finally, we can compute the lower bound for the degradation constant $\alpha$, according to  equation (\ref{eqn:alpha}). 
For the thresholds  $\theta_+ = 3/4$ and $\theta_- = 1/4$,  we have $\alpha \geq \frac{3.4012}{\delta}$, while for $\theta_{C^-} = 1/3$ and $\theta_{C^+} = 2/3$, we have $\alpha \geq \frac{3.6889}{\delta}$.

 %%%%  NOT    %%%%%
\subsubsection*{NOT gate.}
The differential equations for the NOT gate are 
$$
 \begin{cases}
 \frac {dx_{D}}{dt}= \alpha_D K  %\frac{\alpha_{D}}{1+(\frac{x_{B}}{K_{BD}})^{n}}
  - \alpha_{D} \cdot x_{D},\\ 
x_{D}(0)=x_{D_{0}},
\end{cases}
$$
where  $$K= \frac{1}{1+(\frac{x_{B}}{K_{BD}})^{n}}$$

We fix the output concentration thresholds  $\theta_{D^{+}}$, $\theta_{D^{-}}$ and the constant $p>0$, and consider separately the two STL conditions. $\tilde{\theta}_{D^{\pm}}$ are defined as in the previous section.
\begin{enumerate}
%%%%%  1  NOT
\item Case 1 ($x_{B}=\theta_{B^{-}}, x_{D}(0)=0$): Here we need to enforce the condition
$
K  \geqslant \tilde{\theta}_{D^{+}},
$
which results in
$$
 K_{BD}^{n}  \geqslant \frac{\tilde{\theta}_{D^{+}} \theta_{B^{-}}^{n}}{1 - \tilde{\theta}_{D^{+}}}.
$$
%%%%%  2  NOT
\item Case 2 ($x_{B}=\theta_{B^{+}}, x_{D}(0)=x_{D_{0}}$): The condition
$
 K  \leqslant \tilde{\theta}_{D^{-}},
$ gives us
$$
 K_{BD}^{n}  \leqslant \frac{\tilde{\theta}_{D^{-}} \theta_{B^{+}}^{n}}{1 - \tilde{\theta}_{D^{-}}},
$$
\end{enumerate}
Taking the intersection, and imposing that the resulting set is non-void, we get
%Solution iff $ (\frac{\theta_{B^{+}}}{\theta_{B^{-}}})^{n}\ \geqslant \frac{\tilde{\theta}_{D^{+}}}{\tilde{\theta}_{D^{-}} }   \frac{1 - \tilde{\theta}_{D^{-}}}{1 - \tilde{\theta}_{D^{+}}}$ then if: 
$$n \geqslant  \frac{1}{\log(\frac{\theta_{B^{+}}}{\theta_{B^{-}}})}\log\Big(\frac{\tilde{\theta}_{D^{+}}}{\tilde{\theta}_{D^{-}} }   \frac{1 - \tilde{\theta}_{D^{-}}}{1 - \tilde{\theta}_{D^{+}}}\Big)$$ and
$$
 \theta_{B^{-}}\big( \frac{\tilde{\theta}_{D^{+}} }{1 - \tilde{\theta}_{D^{+}}}\big)^{\frac{1}{n}}  \leqslant K_{BD}  \leqslant \theta_{B^{+}} \big( \frac{\tilde{\theta}_{D^{-}} }{1 - \tilde{\theta}_{D^{-}}}\big)^{\frac{1}{n}}.
$$

%%%%%%   OR    %%%%%%
\subsubsection*{OR gate.}
The ODE systems for the OR gate is:
$$
 \begin{cases}
\frac {dx_{S}}{dt}=\alpha_{S} \cdot  \frac{(\frac{x_{E}}{K_{ES}})^{n}+ (\frac{x_{G}}{K_{GS}})^{n}}{1+(\frac{x_{E}}{K_{ES}})^{n}+(\frac{x_{G}}{K_{GS}})^{n}}- \alpha_{S} \cdot x_{S}, \\
x_{S}(0)=x_{S_0},
\end{cases}
$$
with $K$ now defined as
$$K= \frac{(\frac{x_{E}}{K_{ES}})^{n}+ (\frac{x_{G}}{K_{GS}})^{n}}{1+(\frac{x_{E}}{K_{ES}})^{n}+(\frac{x_{G}}{K_{GS}})^{n}}.$$
We can obtain the constraints for the parameters using an approach similar to the one of the AND gate, for a fixed set of activation thresholds  $\theta_{S^-}$ and $\theta_{S^+}$. 
Note that $$K=\theta \mbox{ iff }  (\frac{x_{E}}{K_{ES}})^{n}+ (\frac{x_{G}}{K_{GS}})^{n} = \frac{\theta}{1-\theta} $$

We have  two possible methods also in this case, one stricter, giving an hyperbox, and one less strict, resulting in a curved region.

Remember that, due to Proposition \ref{th:worstCase},  if the input $x_{J}$ is low and the output $x_{O}$ is high then the worst-case input signal is $\hat{x}_{J}=0$; Hence, the analytic treatment of the corresponding cases is very simple. 

\paragraph{Method 1.}
For the parameters $K_{ES}$ and $K_{GS}$ we obtain the following  bounds:
$$   $$
$$   \theta_{E^-} \big( \frac{2-2 \tilde{\theta}_{S^{-}}}{\tilde{\theta}_{S^{-}}} \big)^{1/n}  \leqslant K_{ES} \leqslant \theta_{E^+} \big( \frac{1-\tilde{\theta}_{S^{+}}}{\tilde{\theta}_{S^{+}}} \big)^{1/n} $$
$$   \theta_{G^-} \big( \frac{2-2\tilde{\theta}_{S^{-}}}{\tilde{\theta}_{S^{-}}} \big)^{1/n}    \leqslant K_{GS} \leqslant \theta_{G^+} \big( \frac{1-\tilde{\theta}_{S^{+}}}{\tilde{\theta}_{S^{+}}} \big)^{1/n},    $$
resulting in the following constraint on $n$:
$$n \geq \frac{1}{\min\{ \log(\frac{\theta_{G^{+}}}{\theta_{G^{-}}}),\log(\frac{\theta_{E^{+}}}{\theta_{E^{-}}})\}}  
\log\left(\frac{\tilde{\theta}_{S^{+}}}{\tilde{\theta}_{{S^{-}}}}  \cdot \frac{2 - 2 \tilde{\theta}_{{S^{-}}}}{1 - \tilde{\theta}_{S^{+}}}\right). $$ 

\paragraph{Method 2.} A more refined analysis gives us the following set of parameters: 

\begin{eqnarray*}
& \Big\{  \theta_{E^{-}} \Big(\frac{1-\tilde{\theta}_{S^{-}}}{\tilde{\theta}_{S^{-}}}\Big)^{\frac{1}{n}} <  K_{ES} \leqslant  \theta_{E^{+}} \Big(\frac{1-\tilde{\theta}_{S^{+}}}{\tilde{\theta}_{S^{+}}}\Big)^{\frac{1}{n}},  & \\
& \theta_{G^{-}} \Big(\frac{1-\tilde{\theta}_{S^{-}}}{\tilde{\theta}_{S^{-}}} \Big)^{\frac{1}{n}} < K_{GS} \leqslant \theta_{G^{+}} \Big(\frac{1-\tilde{\theta}_{S^{+}}}{\tilde{\theta}_{S^{+}}}\Big)^{\frac{1}{n}},  & \\
& K_{ES}\geqslant \theta_{E^{-}} \Big( \frac{1}{\frac{\tilde{\theta}_{{S}^{-}}}{1-\tilde{\theta}_{{S}^{-}}}-\frac{\theta^{n}_{{G}^{-}}}{K_{GS}^{n}}} \Big)^{\frac{1}{n}} \Big\}  & \\
\end{eqnarray*}

\begin{figure}
\centering
\includegraphics[width=0.9\textwidth]{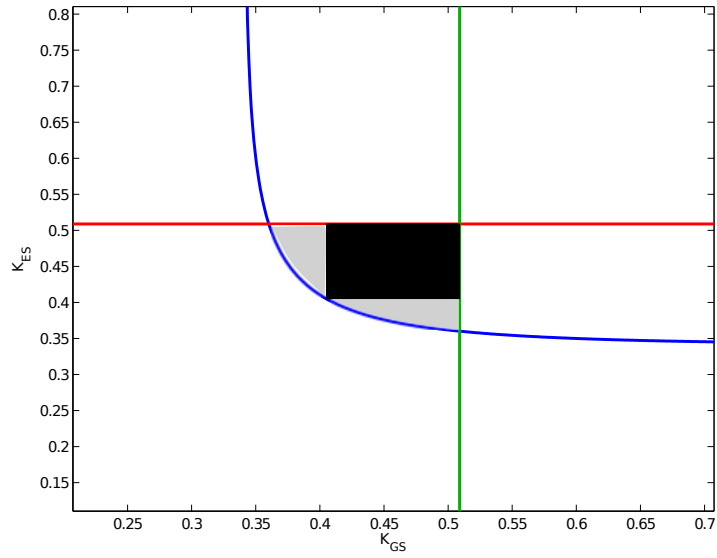}
\caption{The grey region inside the intersection of the curves $
 K_{ES}= \theta_{E^{-}} \Big( \frac{1}{\frac{\tilde{\theta}_{{S}^{-}}}{1-\tilde{\theta}_{{S}^{-}}}-\frac{\theta^{n}_{{G}^{-}}}{K_{GS}^{n}}} \Big)^{\frac{1}{n}}  
$(in {\color{blue}blue}),
$
 K_{GS}= \theta_{G^{+}} \Big(\frac{1-\tilde{\theta}_{S^{+}}}{\tilde{\theta}_{S^{+}}}\Big)^{\frac{1}{n}}$ (in {\color{green!60!black}green}) and
$
 K_{ES}= \theta_{E^{+}} \Big(\frac{1-\tilde{\theta}_{S^{+}}}{\tilde{\theta}_{S^{+}}}\Big)^{\frac{1}{n}}$ (in  {\color{red}red})  is the validity domain of the parameters $K_{AC}$ and $K_{BC}$ for   $\theta_+ = 3/4$ and $\theta_- = 1/4$. The black region is the one identified by the first method.}
\label{pOR}
\end{figure}

We also obtain the following constraint on the  parameter $n$:
$$n > \frac{1}{\min\{ \log(\frac{\theta_{G^{+}}}{\theta_{G^{-}}}),\log(\frac{\theta_{E^{+}}}{\theta_{E^{-}}})\}}  
\log\left(\frac{\tilde{\theta}_{S^{+}}}{\tilde{\theta}_{{S^{-}}}}  \cdot \frac{1 - \tilde{\theta}_{{S^{-}}}}{1 - \tilde{\theta}_{S^{+}}}\right). $$
In Figure \ref{pOR}, we compare the two validity sets for $\theta_{S^-}=1/4$ and $\theta_{S^+}=3/4$, $p=0.1$, $n=3$

\end{document}